\documentclass[aps,prd,superscriptaddress,amsfonts,amssymb,amsmath,eqsecnum,nofootinbib,twocolumn,floatfix]{revtex4-1}
  \usepackage{color,graphicx}
  \usepackage[utf8]{inputenc}
  \usepackage[T2A]{fontenc}
  \usepackage[english]{babel}
  \definecolor{darkblue}{rgb}{0,0,0.7}
 \definecolor{darkred}{rgb}{0.7,0,0}
 \definecolor{darkgreen}{rgb}{0,0.4,0}
 \usepackage[unicode, colorlinks, citecolor=darkblue, linkcolor=darkred, urlcolor=blue]{hyperref}

 \allowdisplaybreaks
\begin{document}

\author{Aleksandr A. Movsisian}

\affiliation{Faculty of Physics, M.V. Lomonosov Moscow State University, Leninskie Gory, Moscow 119991, Russia}

\author{Sergey P. Vyatchanin}
\affiliation{Faculty of Physics, M.V. Lomonosov Moscow State University, Leninskie Gory, Moscow 119991, Russia}
\affiliation{Quantum Technology Centre, M.V. Lomonosov Moscow State University, Leninskie Gory, Moscow 119991, Russia}

%

\date{\today}
	
\title{Squeezing for Broadband Multidimensional Variational Measurement}

\begin{abstract}
Broadband multidimensional variational measurement allows to overcome Standard Quantum Limit (SQL) of a classical mechanical force detection, resulting from quantum back action, which perturbs evolution of a mechanical oscillator. In this optomechanic scheme detection of a resonant signal force acting on a linear mechanical oscillator coupled to a system with three optical modes with separation nearly equal to the mechanical frequency. The measurement is performed by optical pumping of the central optical mode and measuring the light escaping the two other modes. Detection of optimal quadrature components of the optical modes  and post processing result in the back action exclusion in a broad frequency band and  surpassing SQL. We show that optical losses inside cavity restrict back action exclusion due to loss noise.
We also analyze how two-photon (nondegenerate) and conventional (degenerate) squeezing improve sensitivity with account optical losses, considering mainly internal  squeezing.
\end{abstract}

\maketitle

\section{Introduction}

Optical transducers are frequently used to observe mechanical motion. These transducers allow to detect displacement, speed, acceleration, and rotation of mechanical systems. Mechanical motion can change phase or amplitude of the probe light. The sensitivity of the measurement can be extremely high, for example, a relative mechanical displacement much smaller than a proton size can be detected. It was demonstrated in gravitational wave detectors \cite{AbbotLRR2020,aLIGO2015,MartynovPRD16,AserneseCQG15, DooleyCQG16,AsoPRD13, AkutsuPTEP2021}, in magnetometers \cite{ForstnerPRL2012, LiOptica2018}, and in torque sensors \cite{WuPRX2014, KimNC2016, AhnNT2020}. 

There are several reasons restricting the fundamental sensitivity of the measurement. One of them is the fundamental thermal fluctuations in mechanical system (Nyquist noise). However, this obstacle can be considerably decreased  if one  measures a variation of the position during time much smaller than the system ring down time \cite{Braginsky68, 92BookBrKh}.

Another limitation comes from the quantum noise of the meter. On one hand, the accuracy of the measurements  is limited because of their fundamental quantum fluctuations, represented by the shot noise for the optical probe wave, in order to decrease it one have to increase pump power. On the other hand, the sensitivity is impacted by the perturbation of the state of the probe mass due to so called ``back action'', which increases with growth of pump power.   In the case of optical meter the mechanical perturbation results from fluctuations of the light pressure force. Interplay between these two factors leads to a so called standard quantum limit (SQL) \cite{Braginsky68, 92BookBrKh} of the sensitivity. 

SQL is a consequence of noncommutativity between the probe noise and the quantum back action noise. In a simple displacement sensor the probe noise is the phase noise of light and the back action noise is the amplitude noise of light (light pressure noise). The signal is contained in the phase of the probe. The relative phase noise decreases with optical power whereas the relative back action noise increases with the power. The optimal measurement sensitivity corresponds to SQL.  It is not possible to measure the amplitude noise and subtract it from the measurement result, because of phase and amplitude quantum fluctuations of the same wave do not commute. 

The SQL of a mechanical force, acting on free test mass, can be surpassed in a transducer supporting opto-mechanical velocity measurement  \cite{90BrKhPLA,00a1BrGoKhThPRD}. The limit also can be overcome using opto-mechanical rigidity \cite{99a1BrKhPLA, 01a1KhPLA}. Preparation of the probe light in a nonclassical squeezed state \cite{LigoNatPh11, LigoNatPhot13, TsePRL19, AsernesePRL19, YapNatPhot20, YuNature20, CripeNat19,19Bachor} as well as detection of a variation of a strongly perturbed optical quadrature \cite{93a1VyMaJETP, 95a1VyZuPLA, 02a1KiLeMaThVyPRD, AsernesePRL2020} curbs the quantum back action and lifts SQL. The SQL can be surpassed with coherent quantum noise cancellation \cite{TsangPRL2010, PolzikAdPh2014, MollerNature2017} as well as compensation using an auxiliary medium with negative nonlinearity \cite{matsko99prl}.
Optimization of the measurement scheme by usage a few optical frequency harmonics as a probe also allows beating the SQL. A dichromatic optical probe may lead to observation of such phenomena as negative radiation pressure \cite{Povinelli05ol,maslov13pra} and optical quadrature-dependent quantum back action evasion \cite{21a1VyNaMaPRA}. 

The first procedure of back action evading (BAE) for mechanical oscillator, proposed about forty years ago \cite{80a1BrThVo, 81a1BrVoKh}, took advantage of short (stroboscopic) measurements of a  mechanical  coordinate separated by a half period of oscillator. At the same time it was proposed to measure not a coordinate but one of quadrature amplitudes of a mechanical oscillator \cite{Thorne1978, 80a1BrThVo} to perform a BAE. Both propositions are equivalent and can be realized with a pulsing pump \cite{81a1BrVoKh, Clerk08, 18a1VyMaJOSA}.

Recently broadband multidimensional variational measurement for mechanical oscillator was proposed \cite{22PRAVyNaMa}. Mechanical oscillator is coupled to a system with three optical modes, which frequencies $\omega_\pm,\ \omega_0$ are separated by the mechanical frequency $\omega_m$: $\omega_\pm =\omega_0\pm \omega_m$. The measurement is performed by optical pumping of the central optical mode $\omega_0$ and measuring the light escaping the two other modes $\omega_\pm$. Detection of optimal quadrature components of the output waves of modes $\omega_\pm$ {\em separately} provides two channel registration. It allows  to detect back action  and to remove it completely from the measured data via post processing. 


In quantum  measurement of free mass position preparation of the probe light in a squeezed state \cite{LigoNatPh11, LigoNatPhot13, TsePRL19, AsernesePRL19, YapNatPhot20, YuNature20, CripeNat19, KorobkoPRL2017, KorobkoLSA2019, GardnerPRD2022} gives possibility to surpass SQL.
In this paper we analyze how sensitivity of broadband multidimensional variational measurement for detection of small signal force, acting on mechanical oscillator, can be improved by squeezing even in presence of optical losses. We consider two possibilities of two photon (nondegenerate)  and conventional (degenerate) {\em internal} squeezing and compare them with each other. We also analyze and compare sensitivity improvement for internal squeezing for the cases of two photon and conventional squeezing.

\begin{figure}
 \includegraphics[width=0.35\textwidth]{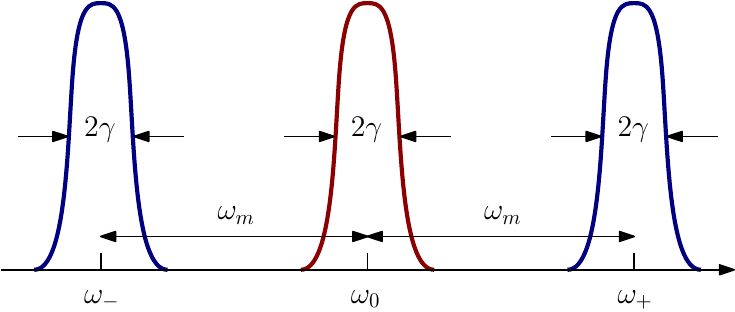}
 \caption{Optomechanical scheme. Three optical modes which frequencies are separated by frequency $\omega_m$ of mechanical oscillator. Optical modes are coupled with the mechanical oscillator. Relaxation rate $\gamma$ is the same for the all three modes, $\gamma\ll \omega_m$. The middle mode with frequency $\omega_0$ is resonantly pumped. }\label{scheme}
\end{figure}

\section{Physical Model}
\label{Model}

\begin{figure}
 \includegraphics[width=0.45\textwidth]{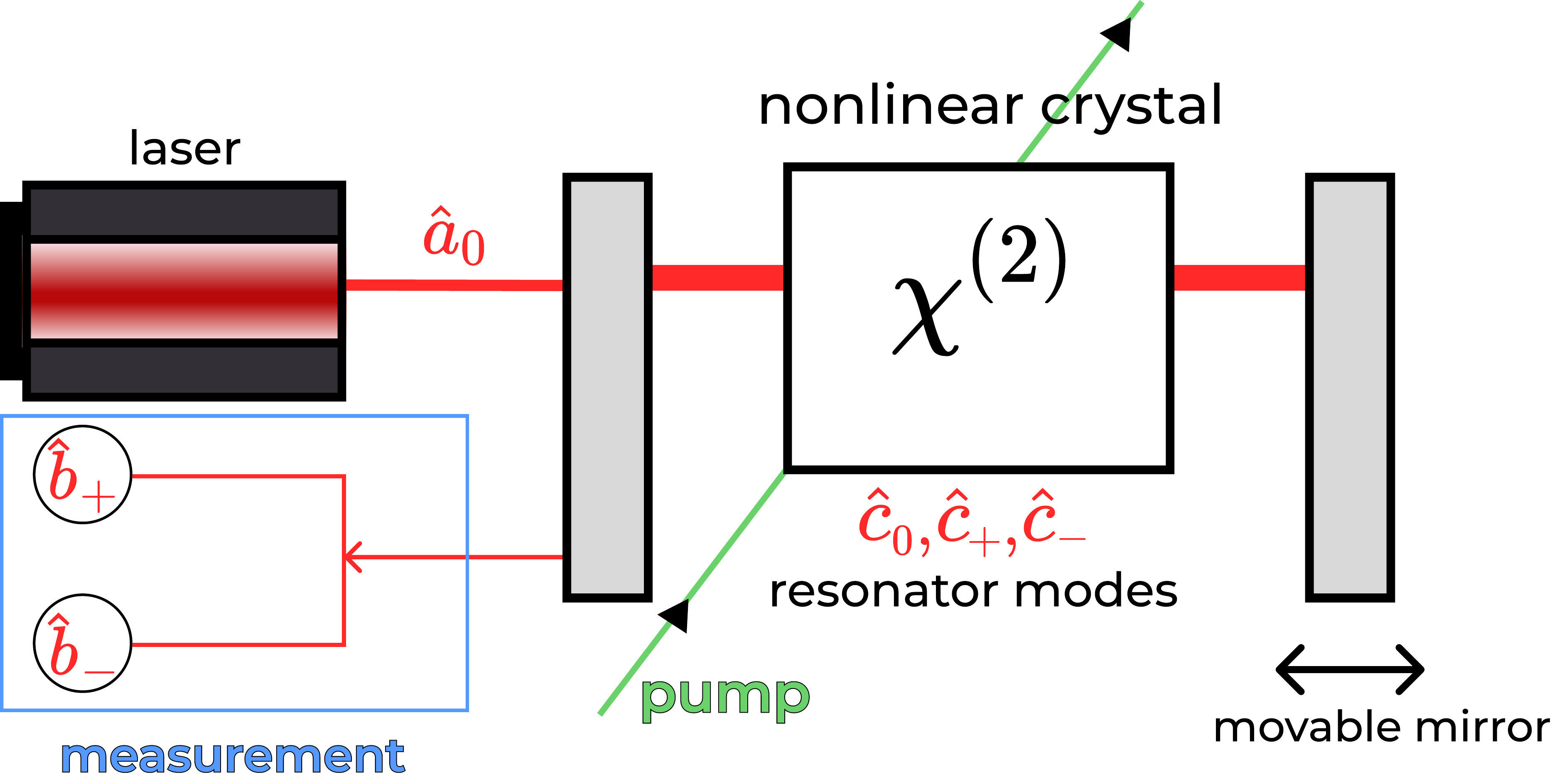}
 \caption{An optomechanical scheme based on a Fabry-Perot resonator with a transmissive front mirror and a non-transmissive movable rear wall. There is a nonlinear crystal inside the resonator, which is pumped at twice the laser carrier frequency to create two photon squeezing in  modes $\omega_\pm$ of the resonator. The output of these modes are then measured separately, details of the measurement are shown in Fig.~\ref{scheme2}. }\label{schemeM}
\end{figure}

Let we have cavity with triplet of optical modes $\omega_-,\ \omega_0,\ \omega_+$ separated from each others by eigen frequency $\omega_m$ of mechanical oscillator as shown on Fig.~\ref{scheme}. The middle mode with frequency $\omega_0$ is resonantly pumped, the modes  $\omega_\pm$ are not pumped. Mass of mechanical oscillator is a movable end mirror, it provides coupling with optical modes. We assume two photon (nondegenerate) {\em internal} squeezing of fields in the modes $\omega_{\pm}$, created by parametric interaction with pump on frequency $2\omega_0$. We detect output of modes $\omega_\pm$.
See scheme on Fig.~\ref{schemeM}.

We  assume that the relaxation rates of the optical modes are identical and characterized with the full width at the half maximum (FWHM) equal to $2\gamma=2(\gamma_0+\gamma_e)$, where $\gamma_0$ characterizes transmittance of input mirror and $\gamma_e$ --- optical losses of cavity's mirrors, relaxation rates $\gamma_0,\ \gamma_e$ are the same for all optical modes.  The  mechanical relaxation rate $\gamma_m$ is small as compared with the optical one. We also assume that frequency synchronization, the conditions of the resolved side band interaction and condition of optical loss smallness are valid:  
\begin{align}
 \label{RSB}
 \omega_\pm =\omega_0\pm \omega_m, \quad  \gamma_m\ll \gamma \ll \omega_m, \quad \gamma_e \ll \gamma_0
\end{align}

\subsection{Hamiltonian}

The generalized Hamiltonian describing the system can be presented in form
\begin{subequations}
   \label{Halt}
  \begin{align}
  H   &= H_0 + H_\text{int}+H_s+ H_{sq} + \\
  & +H_{T\, 0}+H_{\gamma_0} + H_{T\, e}+H_{\gamma_e}+H_{T, \, m}+H_{\gamma_ m} ,\nonumber\\
  \label{H0}
  H_0 &=\hslash \omega_+\hat c_+^\dag \hat c_+  + \hslash \omega_0 \hat c_0^\dag \hat c_0 +\\
    &\qquad + \hslash \omega_-\hat c_-^\dag \hat c_-
         +\hslash \omega_m \hat d^\dag \hat d,\nonumber\\
  \label{Hint}
    H_\text{int} & = \frac{\hslash }{i}
      \left(\eta \left[\hat c_0^\dag \hat c_- + \hat c_+^\dag \hat c_0\right] \hat d -\right.\\
     &\qquad -\left. \eta^* \left[\hat c_0 \hat c_-^\dag+ \hat c_+ \hat c_0^\dag \right]\hat  d^\dag \right),\nonumber\\
   \label{Hs}\hat 
   H_s & = - F_s x_0\left(\hat d+ \hat d^\dag\right),\\
   H_{sq} &= \frac{\hslash \nu}{i}\left( C_{02}\hat c_+^\dag \hat c_-^\dag - C_{02}^*\hat c_+\hat c_-\right). 
  \end{align}
  \end{subequations}
Here $\hslash$ is Plank constant. $H_0$ describes energies of optical modes and mechanical oscillator, $\hat c_0,\ \hat c^\dag_0,\ \hat c_\pm,\ \hat c^\dag_\pm$ are annihilation and creation operators of the corresponding optical modes, $\hat d,\ \hat d^\dag$ are annihilation and creation operators of the mechanical oscillator. The operator of coordinate $\hat x$ of the mechanical oscillator is presented in usual form
\begin{align}
\label{x}
 \hat x= x_0\left(\hat d + \hat d^\dag\right),\quad x_0=\sqrt\frac{\hslash}{2m \omega_m},
\end{align}
where $m$ is mass of oscillator.
$H_\text{int}$ is the Hamiltonian of the interaction between optical and mechanical modes\footnote{
Initially $H_\text{int}\sim (E_0+E_+ +E_-)^2x$, where $E_0,E_-,E_+$ are electric fields of modes $0,-,+$ on surface of mirror (mass of oscillator), it can be transformed into form \eqref{Hint} after omitting fast oscillating terms.}, $\eta\simeq x_0\omega_0/L$ is coupling constant, $L$ is the length of cavity. $H_s$ is a part of Hamiltonian describing  signal force $F_s$. Hamiltonian $H_{sq}$ describes two photon (nondegenerate) internal squeezing, $\nu$ is a constant proportional to nonlinear $\chi^{(2)}$ susceptibility of crystal inside cavity, $C_{02}$ is an expectation value of amplitude of classical pump, acting on frequency $2\omega_0$.  
$H_{T\,0}$ is the Hamiltonian describing the outer environment (regular and fluctuational fields incident on input mirror) and  $H_{\gamma_0}$ is the Hamiltonian of the coupling between the outer environment and the optical modes resulting in decay rate $\gamma_0$; $H_{T\,e}$  and  $H_{\gamma_e}$ describe optical losses. The pump is also included into $H_{\gamma_0}$. Similarly, $H_{T, \, m}$ is the Hamiltonian of the environment and $H_{\gamma_m}$ is the Hamiltonian describing coupling between the environment and the mechanical oscillator resulting in decay rate $\gamma_m$. See Appendix~\ref{IntrDeriv} for details.

\section{Analysis for two photon nondegenerate internal and external squeezing}

We denote the normalized input and output optical amplitudes as $\hat a_{\pm, \,0}$ and  $\hat b_{\pm, \,0}$ correspondingly. Using the Hamiltonian \eqref{Halt} we derive the equations of motion for the intracavity slow amplitudes of fields (see Appendix~\ref{IntrDeriv} for the full derivation).
\begin{subequations}
\begin{align} 
\dot {\hat c}_0+\gamma \hat c_0&=\eta^*\hat c_+ \hat d^\dag - \eta \hat c_- \hat d+\\
 &\qquad  +\sqrt{2 \gamma_0}\,\hat a_0 + \sqrt{2 \gamma_e}\, \hat e_0,\quad \gamma= \gamma_0+\gamma_e\nonumber\\
\dot {\hat c}_++\gamma \hat c_+&=-\eta \hat c_0 \hat d + \nu C_{02}\hat c_-^\dag +\\
& \qquad+\sqrt{2 \gamma_0}\,\hat a_+ +\sqrt{2\gamma_e} \hat e_+, \nonumber\\
\dot { \hat c}_-+\gamma \hat c_-&=\eta^*\hat c_0 \hat d^\dag + \nu C_{02}\hat c_+^\dag+\\
&\qquad + \sqrt{2 \gamma_0}\,\hat a_-+\sqrt{2 \gamma_e}\,\hat e_-,\nonumber\\
\dot {\hat d}+\gamma_m \hat d&=\eta^*(\hat c_0 \hat c_-^\dag + \hat c_0^\dag \hat c_+)  +\sqrt{2 \gamma_m}\,\hat q +f_s.
\end{align}  \label{moveq}
\end{subequations} 
Here $\hat e_\pm$ describe quantum fluctuation on account of optical losses,
Here $\hat q$ is normalized fluctuation force acting on mechanical oscillator, and $f_s$ is normalized signal force (see definition \eqref{fs} below).

The operators $\hat a_\pm,\ \hat e_\pm,\ \hat q$ are characterized with the following commutators and correlators
\begin{subequations}
 \label{commT}	
\begin{align}
	 \label{commA}
	\left[\hat a_\pm(t), \hat a_\pm^\dag(t')\right] &=   \delta(t-t'),\\
	\label{corrA}
	\left\langle\hat a_\pm(t)\, \hat a_\pm^\dag(t')\right\rangle &= \delta(t-t'),\\
	\left[\hat e_\pm(t), \hat e_\pm^\dag(t')\right] &=   \delta(t-t'),\\
	\label{corrE}
	\left\langle\hat e_\pm(t)\, \hat e_\pm^\dag(t')\right\rangle &= \delta(t-t'),\\
	\left[\hat q(t), \hat q^\dag(t')\right] &=   \delta(t-t'),\\
	\label{corrQ}
	\left\langle\hat q(t) \hat q^\dag(t')\right\rangle &= (2n_T +1)\, \delta(t-t'),\\
	& n_T= \frac{1}{ e^{\hslash \omega_m/\kappa_BT} -1 }.
\end{align}
\end{subequations}
Here $\langle \dots \rangle$ stands for ensemble averaging. Correlators \eqref{corrE}, \eqref{corrQ} are true for since the fluctuation fields are considered to be in the vacuum (or thermal for $\hat q$) state. Correlators \eqref{corrA} is true only while incident field is in coherent state, however it should be revised when incident field is in squeezed state. $n_T$ is thermal number of mechanical quanta, $\kappa_B$ is Boltzmann constant, $T$ is the ambient  temperature.

The relation \eqref{corrA} is given for non-squeezed input amplitudes. For {\em external} squeezing it should be revised.

The input-output relations connecting the incident ($\hat a_\pm$) and intracavity ($\hat c_\pm$) amplitudes with output ($\hat b_\pm$) amplitudes are
\begin{align}
 \label{outputT}
  \hat b_\pm= -\hat a_\pm + \sqrt{2\gamma_0} \hat c_\pm.
\end{align} 

It is convenient to separate the expectation values of the wave amplitudes at frequency $\omega_0$ (described by block letters) as well as its fluctuation part (described by small letters) and assume that the fluctuations are small:
 \begin{align}
 \label{expA}
 \hat c_0 & \Rightarrow C_0 +  c_0 ,\quad  
 \end{align}
here $C_0$  stands for the expectation value of the field amplitude in the mode with eigenfrequency $\omega_0$ and  $c_{  0}$ represent the quantum fluctuations of the field in the mode, $|C_0|^2 \gg \langle  c_0^\dag  c_0 \rangle$. Similar expressions can be written for the optical modes with eigenfrequencies  $\omega_{\pm}$ and the mechanical mode  with eigenfrequency $\omega_m$. The normalization of the amplitudes is selected so that $\hslash \omega_0 |A_0|^2$ describes the optical power of incident wave \cite{02a1KiLeMaThVyPRD}. 

We assume in what follows that the expectation amplitudes are real, same as the coupling constant $\eta$, and introduce parametric gain $\kappa$
\begin{subequations}
	\label{real}
\begin{align}
	C_0= C_0^*,\quad A_0=A_0^*,\quad \eta=\eta^*,\\ 
	C_{02}=C_{02}^*, \quad \kappa = \nu C_{02}
\end{align}
\end{subequations}

\begin{figure}
 \includegraphics[width=0.45\textwidth]{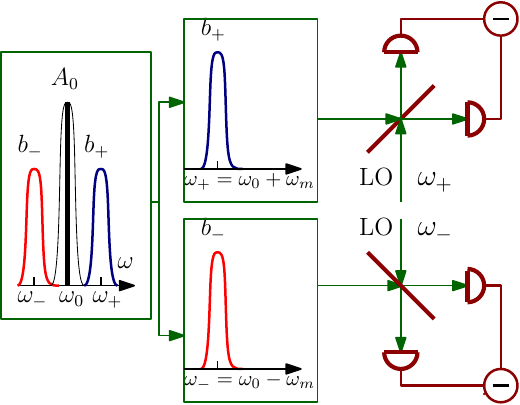}
 \caption{A scheme of the measurement. Quadrature components of the output modes $\omega_\pm$ are measured separately by balanced homodyne detectors with corresponding optimal local oscillators having frequencies $\omega_\pm$. The signal is inferred by processing of the linear combination of the measured results. Essentially, the linear combination in frequency domain should have complex frequency dependent coefficients. }\label{scheme2}
\end{figure}

Substituting  \eqref{expA} and \eqref{real} into the equations of motion \eqref{moveq} and keeping only terms of first order of smallness we  obtain 
\begin{subequations}
	\label{set1}
\begin{align}
\label{c0}
\dot {\hat c}_0+\gamma \hat c_0 &= \sqrt{2 \gamma_0}\hat a_0 +\sqrt{2\gamma_e} e_0,\\
 \dot {\hat c}_++\gamma \hat c_+&=-\eta C_0 \hat d + \kappa\hat c_-^\dag +\\
 & \qquad+\sqrt{2 \gamma_0}\,\hat a_+ +\sqrt{2\gamma_e} \hat e_+, \nonumber\\
 \dot {\hat c}_-+\gamma \hat c_-&=\eta C_0 \hat d^\dag + \kappa\hat  c_+^\dag +\\
 &\qquad + \sqrt{2 \gamma_0}\,\hat a_-+\sqrt{2 \gamma_e}\,\hat e_-,\nonumber\\
 \dot {\hat d}+\gamma_m \hat d&=\eta C_0( \hat c_-^\dag +  \hat c_+)  +\sqrt{2 \gamma_m}\,\hat q +f_s.
\end{align}
\end{subequations}
Outputs $b_\pm$ around frequencies $\omega_\pm$ have to be detected separately, as shown in Fig.~\ref{scheme2}. We also see that fluctuation waves around $\omega_0$ do not influence on field components in the vicinity of frequencies $\omega_\pm$ and the first equation \eqref{c0} separates from the other three, so it is omitted in the further consideration. 

%


The Fourier transform of  operators, for example, $\hat a_\pm$ is defined as follows
\begin{align}
 \label{apmFT}
 \hat a_\pm (t) &= \int_{-\infty}^\infty a_\pm(\Omega) \, e^{-i\Omega t}\, \frac{d\Omega}{2\pi}.
\end{align}
For operators $a_\pm(\Omega)$ the following commutators and correlators are valid:
\begin{align}
 \label{comm1}
  \left[a_\pm(\Omega), a_\pm^\dag(\Omega')\right] &= 2\pi\, \delta(\Omega-\Omega'),\\
  \label{corr1}
  \left\langle a_\pm(\Omega) a_\pm^\dag(\Omega')\right\rangle &= 2\pi\, \delta(\Omega-\Omega')
\end{align}

Similar expressions can be written for the other noise operators ($\hat e_\pm,\ \hat q$). 

We assume that the signal force is a resonant square pulse acting during time interval $\tau$ ($\omega_m\tau\gg 1$):
\begin{align}
\label{Fs}
F_S(t)&= F_{s0}\sin(\omega_m t + \psi_f) = \\
    = & i \left(F_{s}(t) e^{-i\omega_m t} - F_{s}^*(t) e^{i\omega_m t}\right),\quad 
  -\frac{\tau}{2} < t < \frac{\tau}{2},\nonumber\\
  \label{fs}
  f_s(\Omega)& = \frac{F_s(\Omega)}{\sqrt{2\hslash \omega_m m}},\\
  f_{s0}(\Omega)& = \frac{F_{s0}(\Omega)}{\sqrt{2\hslash \omega_m m}}= 2f_s(\Omega).
\end{align}
where  $F_s(\Omega)\ne F_s^*(-\Omega)$ is the Fourier amplitude of $F_s(t)$.  

Let introduce quadrature amplitudes of amplitude and phase
\begin{subequations}
\label{quadDef}
 \begin{align}
  a_{\pm a} &= \frac{a_\pm (\Omega) +a_\pm ^\dag(-\Omega)}{\sqrt 2}\,,\\
	 a_{\pm \phi} &= \frac{a_\pm (\Omega) -a_\pm ^\dag(-\Omega)}{i\sqrt 2}\,.
 \end{align}
\end{subequations}
Analysis shows that sum $(c_{+a} +c_{-a})$ does not contain information on the mechanical motion but contains  the back action term. At the same time difference $(c_{+\phi} -c_{-\phi})$ does not contain information on mechanical motion (term $\sim d_\phi$), but is responsible on back action (see detailes in Appandix \ref{appB}). So it should be useful to introduce sum and difference of the quadratures
\begin{subequations}
\label{SumDif}
\begin{align}
 \label{gDef}
 g_{a\pm} &= \frac{c_{+a}\pm c_{-a}}{\sqrt 2},\quad 
  g_{\phi\pm} = \frac{c_{+\phi}\pm c_{-\phi}}{\sqrt 2},\\
 \label{alphaDef}
 \alpha_{a\pm}&= \frac{a_{+a}\pm a_{-a}}{\sqrt 2}\,,\quad
    \alpha_{\phi\pm}= \frac{a_{+\phi}\pm a_{-\phi}}{\sqrt 2},\\
 \label{betaDef}
 \beta_{a\pm}&= \frac{b_{+a}\pm b_{-a}}{\sqrt 2}\,,\quad
    \beta_{\phi\pm}= \frac{b_{+\phi}\pm b_{-\phi}}{\sqrt 2},\\
  \label{epsDef}
  \epsilon_{a\pm}&= \frac{e_{+a}\pm e_{-a}}{\sqrt 2}\,,\quad
  \epsilon_{\phi\pm}= \frac{e_{+\phi}\pm e_{-\phi}}{\sqrt 2}
\end{align}    
\end{subequations}
Recall we can measure output quadratures from modes $\omega_\pm$ separately, so it is naturally to take sums and differences of quadratures.

\subsection{Measurement of amplitude quadratures}\label{MeasA}

Recall we measure output quadratures of modes $\omega_\pm$ {\em separately} using balanced homodyne detectors with local oscillators frequencies $\omega_\pm$. Let we find output amplitude quadratures. It is convenient to present the solution of set (\ref{ga+}, \ref{ga-}, \ref{da2}) in Appendix \ref{appB} for sum and difference $\beta_{a\pm}$ of the output amplitude quadratures:
\begin{subequations}
 \label{betaMSIa}
 \begin{align}
 	\label{beta+a}
  \beta_{a+}   &= \xi_+\left\{ \alpha_{a+}+ \frac{\mu_+}{\xi_+} \epsilon_{a+}\right\},\\
  \label{xipm}
  \xi_+ =& \frac{\gamma_0-\gamma_e +\kappa+i\Omega }{\gamma_0+\gamma_e -\kappa-i\Omega},\ 
  \xi_-=\frac{\gamma_0-\gamma_e -\kappa+i\Omega }{\gamma_0+\gamma_e +\kappa-i\Omega}\\ 
  \label{mu}
  &\mu_\pm=\frac{2\sqrt{\gamma_0\gamma_e} }{\gamma_0+\gamma_e \mp\kappa-i\Omega},\\
  \label{beta-a}
  \beta_{a-}  &=\xi_-\left[\alpha_{a-} +\frac{\mu_-}{\xi_-}\epsilon_{a-}  -\right. \\
  \label{beta-a2}
   &\qquad \left.- \frac{\mathcal K}{ \gamma_m-i\Omega}\left\{ \alpha_{a+} +\sqrt\frac{\gamma_e}{\gamma_0}\,\epsilon_{a+}\right\} \right]-\\
   \label{beta-a3}
    &\qquad  - \frac{\sqrt{ \xi_-\mathcal K }}{\gamma_m-i\Omega}
            \left(\sqrt {2 \gamma_m} q_a + f_{s\,a}\right),\\
 \label{mathcalK}
        &\quad  \mathcal K\equiv \frac{4  \gamma_0\,\eta^2 C_0^2}{\gamma_0^2-(\kappa +\gamma_e -i\Omega)^2}
 \end{align}
 \end{subequations}
 It is useful to write dimensionless power $\mathcal K$ in form
 \begin{align}
  \label{K}
  \mathcal K &= \mathcal K_0 \,\frac{\gamma(\gamma_0 -\gamma_e) }{\gamma_0^2-(\kappa +\gamma_e -i\Omega)^2},\\ 
  \label{K0}
  \mathcal K_0 &= \frac{4\gamma_0\omega_0 P_{in}}{m\omega_m L^2\gamma^2(\gamma_0 -\gamma_e)},
 \end{align}
where $L$ is cavity length, $P_{in}$ --- input light power.
 
The result of parametric (internal) gain $\kappa$ is two-photon (nondegenerate) unsqueezing of sum quadrature $\beta_{a+}$ \eqref{beta+a}, whereas part \eqref{beta-a} of $\beta_{a-}$ is squeezed. It is obvious if put $\Omega=0$ and account \eqref{RSB}. The  squeezing increases as $\kappa$ grows, but parametric gain is restricted by condition of stability.
\begin{equation}
	\label{maxkappa}
	\kappa \le \gamma_0+\gamma_e
\end{equation}

The two photon (nondegenerate) {\em external} squeezing is described by squeezing (or unsqueezing) of quadratures $\alpha_\pm$ of input field. 
 
As expected, in Eq.~\eqref{beta-a2} the back action term is proportional to the normalized mean energy $\mathcal K$, stored in cavity. However, this term can be excluded by the post processing. One  can measure {\em both}  $\beta_{a+}$ and $\beta_{a-}$ simultaneously and (after proper filtration) subtract term in figure brackets in $\beta_{a+}$ \eqref{beta+a} from term in figure brackets in $\beta_{a-}$ \eqref{beta-a2} to remove the back action. However we can not subtract back action completely, we can subtract term $\sim \alpha_{a+}$ completely but term $\sim \epsilon_{a+}$ survives. It means that we can measure combination 
\begin{subequations}
 \label{beta-acomb}
 \begin{align}
  \tilde \beta_{a-} &= \beta_{a-} +  \frac{\xi_-\mathcal K}{\gamma_m-i\Omega}\,\frac{\beta_{a+}}{\xi_+} =\\
  \label{beta-acomb2}
         &=\xi_-\left[\alpha_{a-} +\frac{\mu_-}{\xi_-}\epsilon_{a-}  + \right. \\
   \label{beta-acomb3}
         &\qquad \left.+ \frac{\mathcal K}{ \gamma_m-i\Omega}\left\{  \sqrt\frac{\gamma_e}{\gamma_0}\,\frac{\epsilon_{a+}}{\xi_+^*}\right\} \right]-\\
   \label{beta-acomb4}
         &\qquad  - \frac{\sqrt{ \xi_-\mathcal K }}{\gamma_m-i\Omega}
         \left(\sqrt {2 \gamma_m} q_a + f_{s\,a}\right)
 \end{align}
\end{subequations}
 %

Essentially, the coefficient needed for suppression of the back action is complex, it depends on the spectral frequency $\Omega$. It means post processing filtration.

The term \eqref{beta-acomb3} describes residual back action and it is loss fluctuations $\epsilon_{a+}$ are responsible for it. Obviously that residual back action restricts sensitivity but due to smallness of optical loss ($\gamma_e\ll \gamma_0$) SQL can be surpassed.  
 
We find the force detection condition using  single-sided power spectral density $S_{f}(\Omega)$ for signal force \eqref{Fs}. Assuming that the detection limit corresponds to the signal-to-noise ratio exceeding unity we obtain recalculating  \eqref{beta-a} to signal force:
\begin{subequations}
 \label{tildebeta-a}
 \begin{align}
  \tilde \beta_{a-}& \frac{(\gamma_m-i\Omega)}{\sqrt{\xi_-\mathcal K}}= -
  f_{s0} - \sqrt{2\gamma_m}\, q_a +\\
  &\quad +\frac{\sqrt \xi_-\,(\gamma_m-i\Omega)}{\sqrt{\mathcal K}}\left(\alpha_{a-} +\frac{\mu_-}{\xi_-}\epsilon_{a-}\right) -\\
  &\quad  -  \sqrt{\xi_-\mathcal K}\left\{  \sqrt\frac{\gamma_e}{\gamma_0}\,\frac{\epsilon_{a+}}{\xi_+^*}\right\}\\
  \label{SfDef}
  f_{s0} &\ge \sqrt{S_{f}(\Omega)\cdot \frac{\Delta\Omega}{2\pi} } ,
 \end{align}
\end{subequations}
 where $\Delta \Omega \simeq 2\pi/\tau$ is defined by time $\tau$ of signal force action. 
 
 \subsubsection{No squeezing, no optical losses, no subtraction of back action}
 
\begin{figure}
 \includegraphics[width=0.45\textwidth]{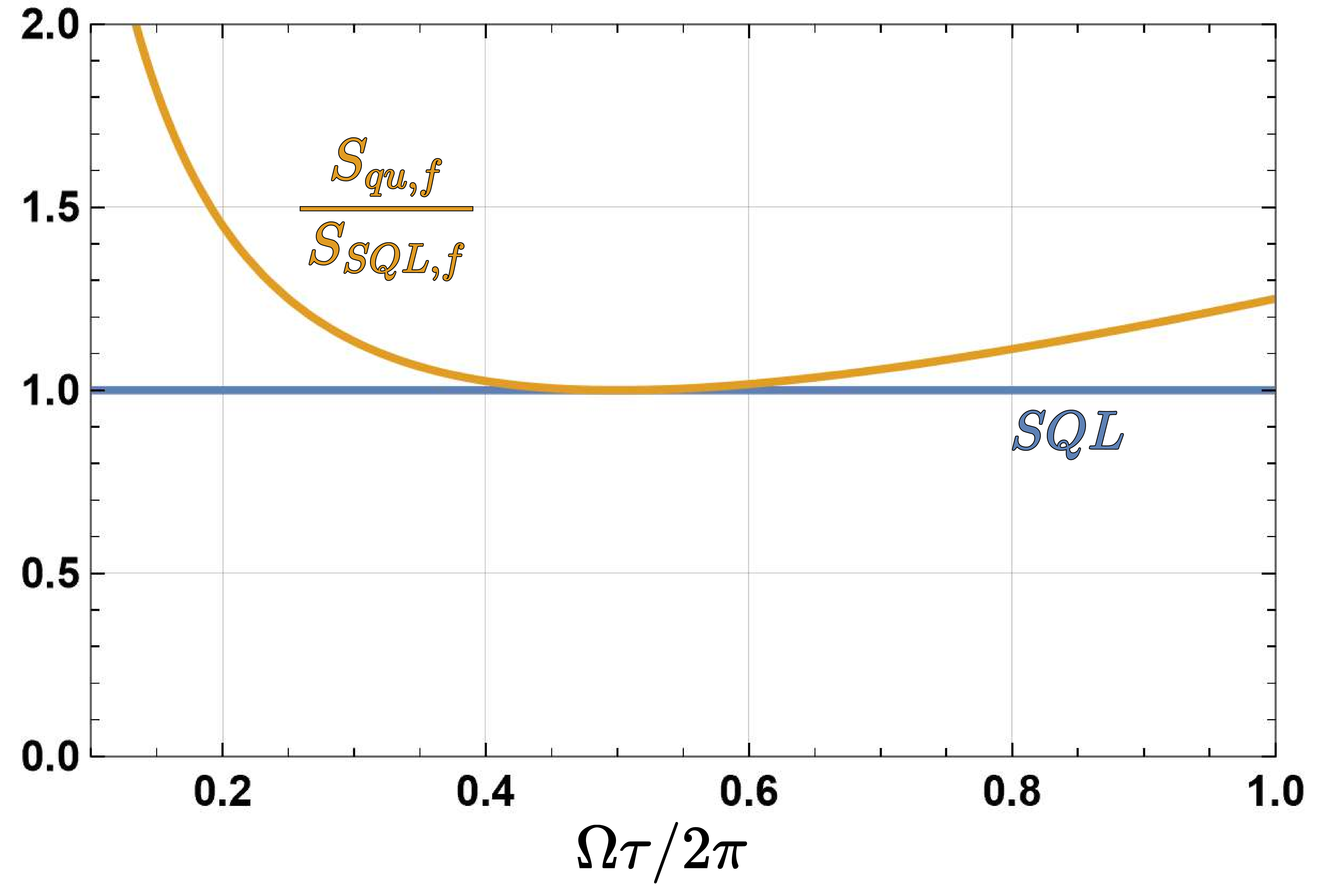}
 \caption{Plot of spectral density $S_{qu,f}$ \eqref{SQu} and $S_{SQL,f}$ \eqref{SSQL} as function of spectral frequency $\Omega$, dimensionless power $\mathcal K_0= \pi/\tau$ \eqref{K0}, $\tau$ is time of signal force action which is chosen to be $\tau=0.28$~msec. }\label{FigSQL}
\end{figure}

 Let consider simplest case without squeezing ($\kappa=0$) and losses ($\gamma_e=0$),  measurement of $ \beta_{a-}$ (\ref{beta-a} -- \ref{beta-a3}) {\em only} (no subtraction of back action).  Using (\ref{corr1e}, \ref{corrq}) we derive 
 \begin{align}
 \gamma_e&=0,\quad \kappa=0,\quad \Rightarrow\quad |\xi_\pm|=1,\ |\mu_\pm|=0\\
  \label{Sf}
  S_{f}^{\beta_{a-}}(\Omega) &= 2\gamma_m\big(2n_T+1\big) + S_{qu,f},\\
    S_{qu,f} & \ge S_{SQL,f},\\ 
  \label{SQu}
  S_{qu,f} &=  \frac{\gamma_m^2+\Omega^2}{\mathcal K} +\mathcal K ,\\
  S_{SQL,f} &= 2\sqrt{\gamma_m^2+\Omega^2} 
  \label{SSQL}
 \end{align}
The sensitivity is restricted by SQL.  Here we assume that normalized pump $\mathcal K$ \eqref{K} practically does not depend on spectral frequency $\Omega$ due to condition \eqref{RSB} and $\Omega\ll \gamma_0$. Strictly speaking, formula \eqref{SSQL} is a kind of mind game --- minimization is taken at optimal $\mathcal K$, depending of spectral frequency $\Omega$, it can not be realized in experiment.  However, more accurate consideration gives result only slightly differing from result with usage \eqref{SSQL} (see details in Appendix \ref{appSQL}). So below we use \eqref{SSQL} for SQL characterization. The examples of plots of $S_{qu,f}$ and $S_{SQL,f}$ are presented on Fig~\ref{FigSQL}, where we put $\mathcal K_0=\pi/\tau$ \eqref{K0}, $\tau$ is time of signal force action. For parameters listed in Table~\ref{table1} it corresponds to $P_{in}\simeq 10^{-2}$~W.

The thermal noise (term $\sim \gamma_m$ in \eqref{Sf} prevents detection of signal force in any opto-mechanical device. 
Below we assume that Braginsky condition \cite{Braginsky68, 92BookBrKh} of smallness of thermal noise as compared with SQL
\begin{align}
\label{BrCond}
    B= \frac{n_T \omega_m\tau}{Q}\ll 1
\end{align}
is fulfilled ($Q$ is a mechanical quality factor). The main requirement for this is large ring down time $\gamma_m^{-1}$ and fast interrogation time $\tau$, i.e. $\gamma_m\tau\ll 1$. Note, for parameters listed in Table~\ref{table1} factor $B\simeq 0.75$ and only for $Q\simeq 10^9$ condition \eqref{BrCond} is realized.
\begin{table}
 \caption{Parameters of a mechanical oscillator (SiN membrane) and optical cavity, used for estimates.} \label{table1}
 \begin{tabular}{||c | c | c||}
 \hline
  \multicolumn{3}{||c||}{Membrane} \\
 \hline
 Mass, $m$ & 50 & $10^{-9}$ g\\
 Frequency, $\omega_m/2\pi$& 350 & $10^3$ Hz\\
 Quality factor $Q=\frac{\omega_m}{2\gamma_m}$ & $10^8$ & \\
 Temperature, $T$ & 20 & K$^\circ$ \\
 Thermal phonons number, $n_T$ & $1.2\cdot 10^6$ &\\
 Time of signal force $\tau=10 \cdot\frac{2\pi}{\omega_m}$ & $28\cdot 10^{-6}$ & sec \\
 \hline
  \multicolumn{3}{||c||}{Cavity} \\
 \hline
 Length of cavity & 10 & cm \\
 Power transmittance of input mirror, $T^2$& $3\cdot 10^{-4}$ &\\
 Power losses (in cavity), $\epsilon^2$ & $10^{-6}$ &  \\
 Bandwidth , $(\gamma_0 + \gamma_e)/2\pi$ & $36$ & $10^{3}$ Hz\\
 Wave length, $\lambda= 2\pi c/\omega_0$ & 1.55 & $10^{-6}$ m \\ 
 Input power $P_{in}$ & 10 & $10^{-3}$ W \\
 \hline
 \end{tabular}
\end{table}

\subsubsection{No squeezing, no optical losses, but subtraction of back action}

We again consider simplest case without squeezing and losses ($\kappa=\gamma_e=0$), but with subtraction of back action via post processing. It means the measurement of combination $\tilde\beta_{a-}$ \eqref{beta-acomb}. Then  the spectral density is {\em not} limited by SQL:
 \begin{align}
 \label{324}
 	\gamma_e&=0,\quad \kappa=0,\quad \Rightarrow\quad |\xi_\pm|=1,\ |\mu_\pm|=0\\
  \label{Sfb}
  S_{f}^{\tilde \beta_{a-}}(\Omega) &= 2\gamma_m\big(2n_T+1\big) + \frac{\gamma_m^2+\Omega^2}{\mathcal K}
 \end{align}
Here the first term describes thermal noise and the second one stands for the quantum measurement noise (shot noise) decreasing with the power increase. The back action term is excluded {\em completely} \cite{22PRAVyNaMa}. 

\begin{figure}
 \includegraphics[width=0.45\textwidth]{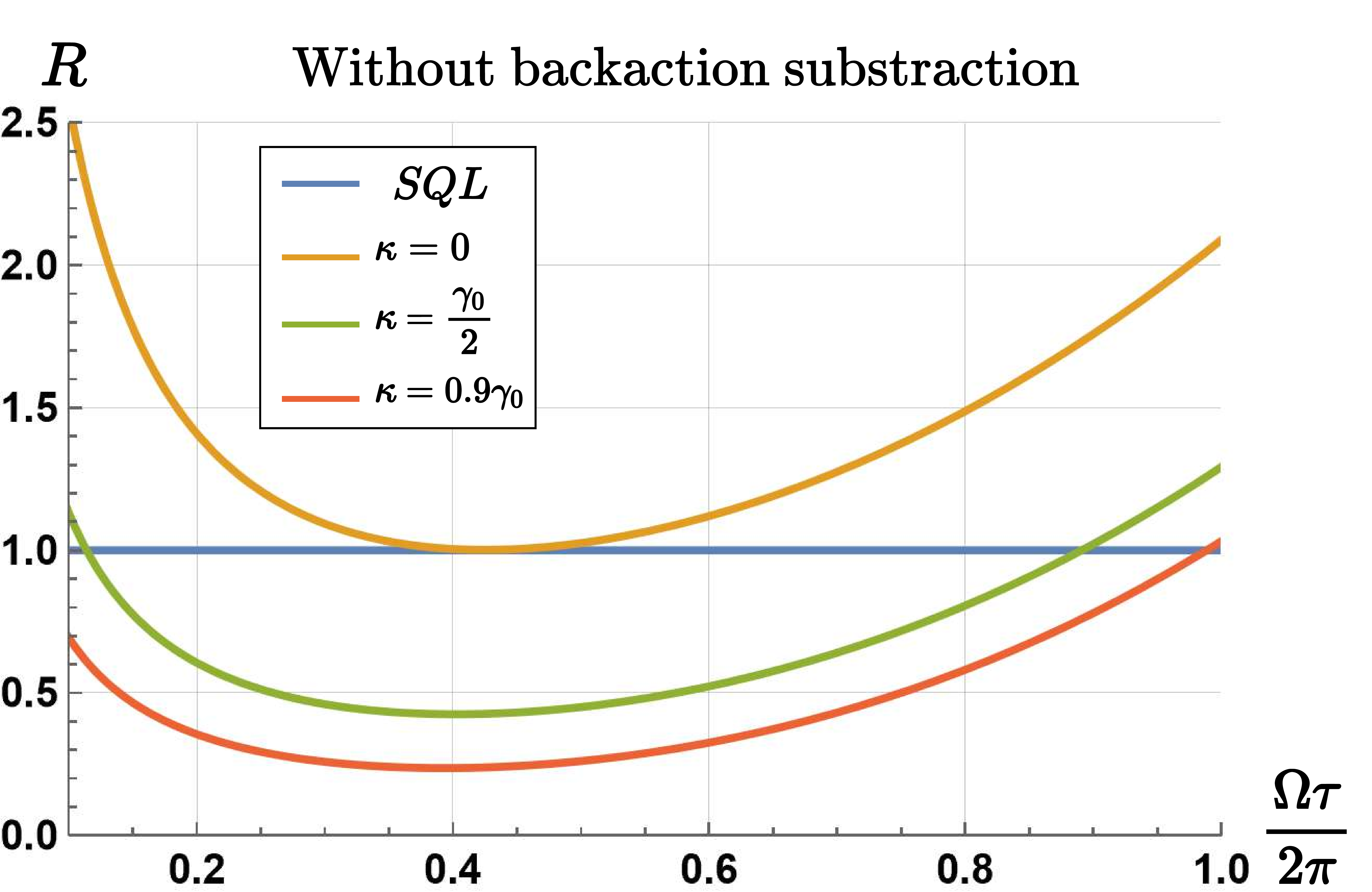}
 \caption{Plots of ratio $R=S_{f}^{\beta_{a-}}/S_{SQL,f}$ of spectral densities (\ref{Sfc}, \ref{SSQL}) as function of spectral frequency $\Omega$ at condition \eqref{condSfc} and $\gamma_m=0$. Dimensionless power $\mathcal K_0= \pi/\tau$ \eqref{K0}, $\tau$ is time of signal force action. Other parameters are taken from Table~\ref{table1}.}\label{WithBA}
\end{figure}

\subsubsection{No external squeezing, nondegenerate squeezing inside cavity, account of losses, no back action subtraction}\label{SecNoBA}

Here we assume that there are two-photon (nondegenerate) internal squeezing and optical losses ($\kappa\ne 0, \gamma_e\ne 0$), input fluctuations fields ($\alpha_{a\pm},\ \epsilon_\pm$) are in vacuum state (no external squeezing). But back action subtraction is not applied. Using (\ref{beta-a}, \ref{beta-a2}, \ref{beta-a3}) we derive power spectral density for this case
\begin{subequations}
\label{Sfc}
 \begin{align}
  \label{condSfc}
 	\gamma_e&\ne 0,\quad \kappa\ne 0,\quad \text{vacuum input} \ \Rightarrow\\
 	\label{Sfc2}
 	S_{f}^{\beta_{a-}}(\Omega) &= 2\gamma_m\big(2n_T+1\big) +\\
    \label{Sfc3}
 	&\quad  + \frac{\gamma_m^2+\Omega^2}{|\mathcal K|}\left(|\xi_-| +\frac{|\mu_-|^2}{|\xi_-|}\right)+\\
  \label{Sfc4}
 	&\quad + |\xi_-\mathcal K|\left\{1+  \frac{\gamma_e}{\gamma_0}\right\}
 \end{align}	
\end{subequations}
Here first term \eqref{Sfc2} describes thermal noise, second one \eqref{Sfc3} --- measurement error and last term \eqref{Sfc4} --- back action.

We present on Fig.~\ref{WithBA} the plots of spectral density \eqref{Sfc}, normalized to SQL (i.e.  $R=S_{f}^{\beta_{a-}}/S_{SQL,f}$) for different parameters of squeezing. For case without squeezing $\kappa=0$ we have usual SQL, but for squeezing $\kappa =\gamma_0/2,\ 0.9\, \gamma_0$ we have considerable gain of sensitivity. 

Recall, system is stable at condition \eqref{maxkappa}.

\begin{figure}
 \includegraphics[width=0.45\textwidth]{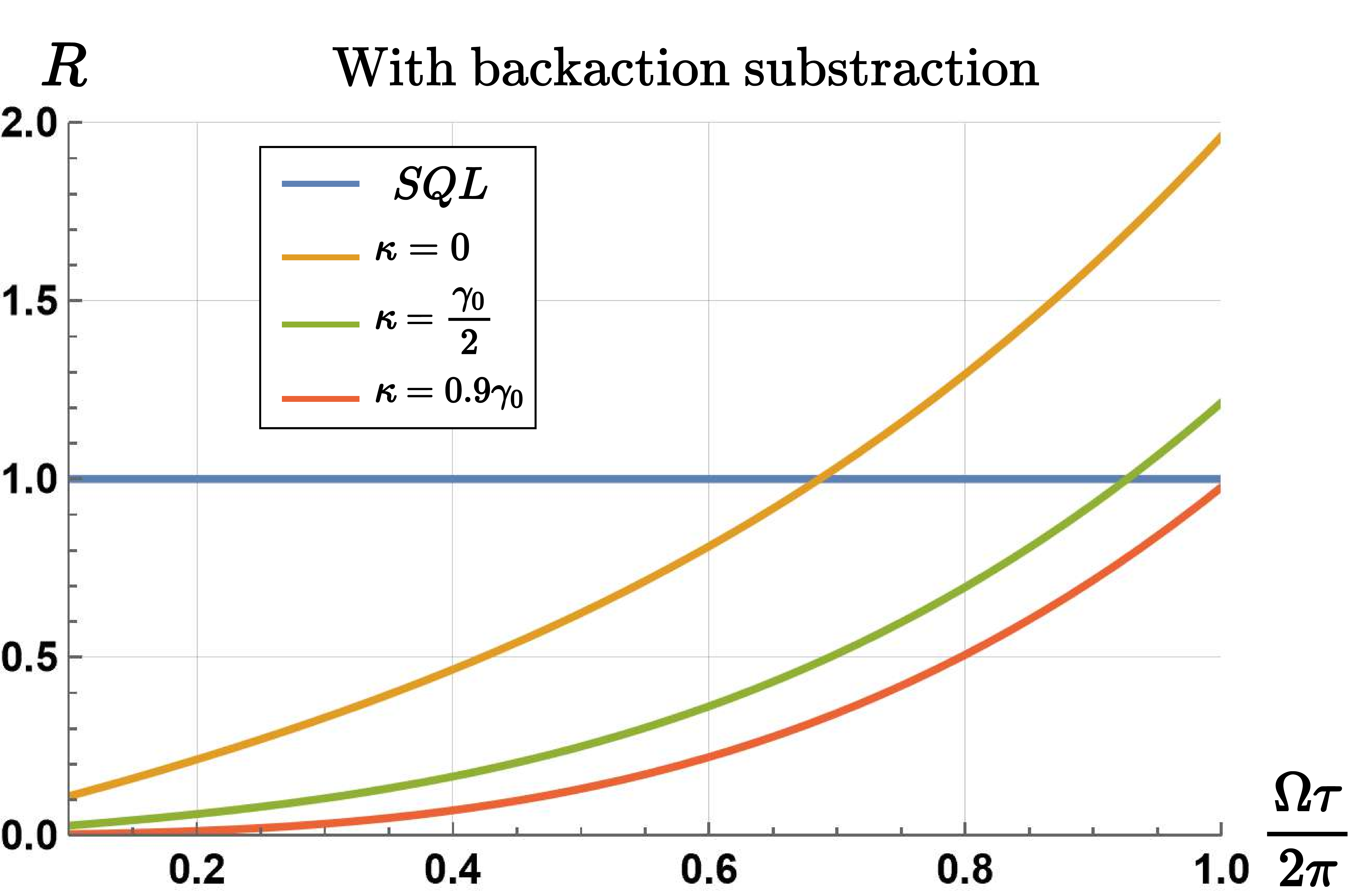}
 \caption{Plots of ratio $R=S_{f}^{\tilde\beta_{a-}}/S_{SQL,f}$ of spectral densities (\ref{Sfd}, \ref{SSQL}) as function of spectral frequency $\Omega$ at condition \eqref{condSfd} and $\gamma_m=0$. Dimensionless power $\mathcal K_0= \pi/\tau$ \eqref{K0}, $\tau$ is time of signal force action. Other parameters are taken from Table~\ref{table1}.}\label{NoBA}
\end{figure}

\begin{figure}
 \includegraphics[width=0.45\textwidth]{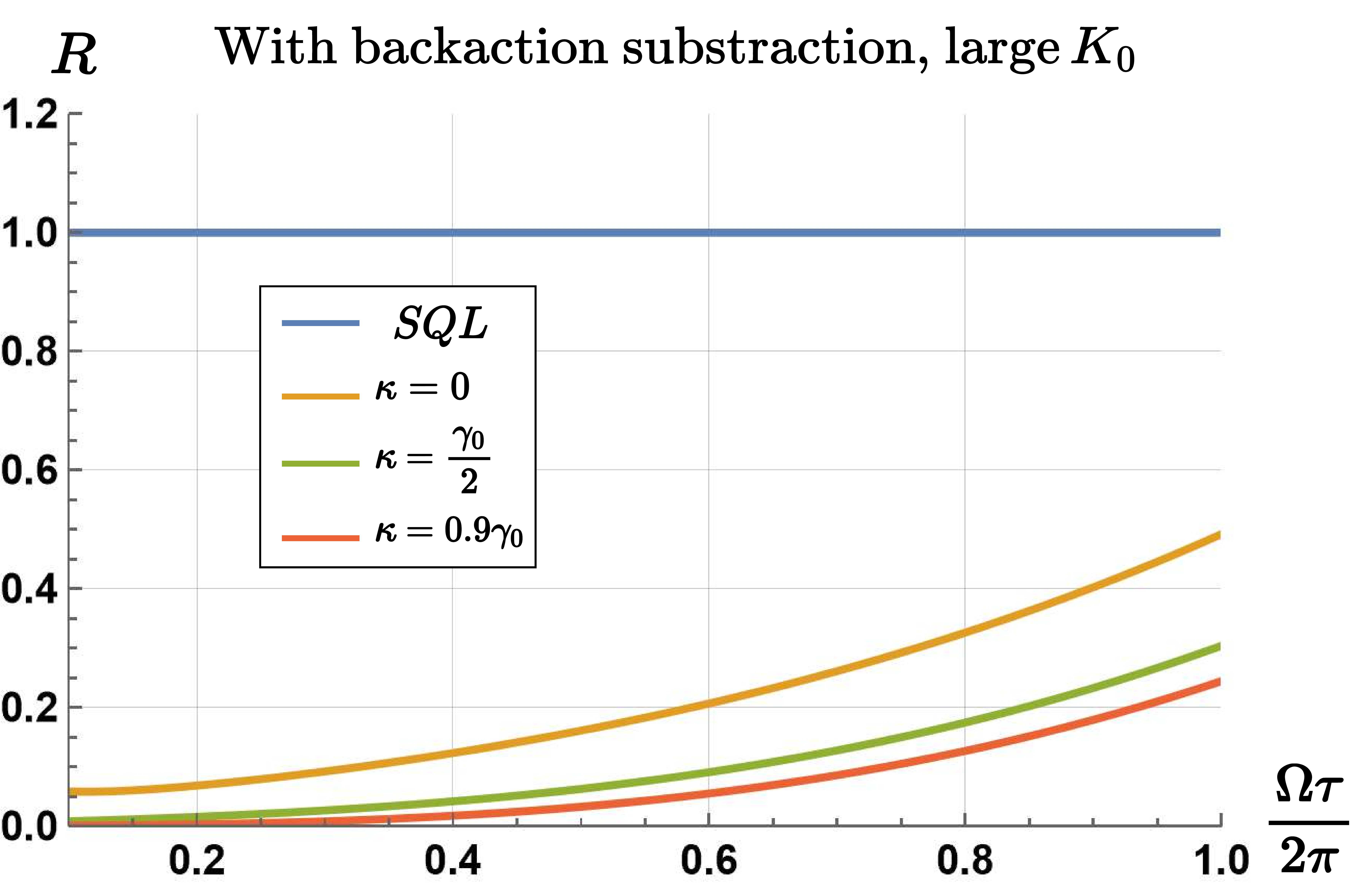}
 \caption{Plots of ratio $R=S_{f}^{\tilde\beta_{a-}}/S_{SQL,f}$ of spectral densities (\ref{Sfd}, \ref{SSQL}) as function of spectral frequency $\Omega$ at condition \eqref{condSfd} and $\gamma_m=0$. In contrast to plots on Fig.~\ref{NoBA} dimensionless power is 4 times larger: $\mathcal K_0= 4\pi/\tau$. Other parameters are the same.}\label{NoBA2}
\end{figure}

\subsubsection{No external squeezing, nondegenerate squeezing inside cavity, account of losses and back action subtraction}

Here we assume the case as in previous subsection \ref{SecNoBA} with the exception of subtraction  back action via post processing. Using \eqref{tildebeta-a} we derive power spectral density for this case
\begin{subequations}
 \label{Sfd}
 \begin{align}
 \label{condSfd}
 	\gamma_e&\ne 0,\quad \kappa\ne 0,\quad \text{vacuum input} \ \Rightarrow\\
 	\label{Sfd2}
 	S_{f}^{\tilde \beta_{a-}}(\Omega) &= 2\gamma_m\big(2n_T+1\big) +\\
 	&\quad  + \frac{\gamma_m^2+\Omega^2}{|\mathcal K|}\left(|\xi_-| +\frac{|\mu_-|^2}{|\xi_-|}\right)+\\
  \label{SfbaRes}
 	&\quad + |\xi_-\mathcal K|\left\{  \frac{\gamma_e}{\gamma_0|\xi_+|^2}\right\}
 \end{align}	
\end{subequations}

Subtraction of back action is possible only partially, term \eqref{SfbaRes} describes residual back action produced by optical loss fluctuations $\sim \epsilon_{a+}$. Corresponding plots are presented on Fig.~\ref{NoBA}. We see that in contrast to Fig.~\ref{WithBA} back action subtraction gives considerable gain on low frequencies.

For plots on Figs.~\ref{FigSQL}, \ref{WithBA}, \ref{NoBA} we used the same normalized power $\mathcal K_0$ \eqref{maxkappa}. For larger $\mathcal K$ the back action subtraction gives more strong SQL surpassing in more wide bandwidth. On Fig.~\ref{NoBA2} plots are presented for $\mathcal K_0=4\pi/\tau$ --- 4 times larger than for plots on Figs.~\ref{FigSQL}, \ref{WithBA}, \ref{NoBA}. It illustrates that even for 4 times larger pump $\mathcal K_0$ provides dramatically increased sensitivity in wider bandwidth.

\subsection{Measurement of phase quadratures} \label{MeasPhi}
 
One can measure sum and differences of the phase quadratures instead of the amplitude quadratures. Solving set (\ref{gphi+}, \ref{gphi-}, \ref{dphi2}) we arrive at
\begin{subequations}
 \label{betaMSIphi}
 \begin{align}
 	\label{beta-phi}
  \beta_{\phi-}   &= \xi_+\left\{ \alpha_{\phi-}+ \frac{\mu_+}{\xi_+} \epsilon_{\phi-}\right\},\\
  \label{beta+phi}
  \beta_{\phi+}  &=\xi_-\left[\alpha_{\phi+} +\frac{\mu_-}{\xi_-}\epsilon_{\phi+}  -\right. \\
  \label{beta+phi2}
   &\qquad \left.- \frac{\mathcal K}{ \gamma_m-i\Omega}\left\{ \alpha_{\phi-} +\sqrt\frac{\gamma_e}{\gamma_0}\,\epsilon_{\phi-}\right\} \right]-\\
   \label{beta+phi3}
    &\qquad  - \frac{\sqrt{ \xi_-\mathcal K }}{\gamma_m-i\Omega}
            \left(\sqrt {2 \gamma_m} q_\phi + f_{s\,\phi}\right).
 \end{align}
 We see that formulas for amplitude output quadratures \eqref{betaMSIa} transform into formulas \eqref{betaMSIphi} just by following substitutions
 \begin{align}
   \beta_{a\pm} &\to \beta_{\phi\mp},\ \alpha_{a\pm} \to \alpha_{\phi\mp},\ d_a\to d_\phi,\ f_{s\,a}\to f_{s\,\phi} 
 \end{align}
 whereas coefficients $\xi_\pm,\ \mu_\pm$ do not change.
 \end{subequations}
So all consideration in previous subsection \ref{MeasA} can be easy rewritten for the case of phase quadratures.  In particular one can measure quadratures $\beta_{\pm\phi}$ independently and by the same way can subtract back action proportional to $\beta_{\phi-}$ from $\beta_{\phi +}$. Back action removal is possible only {\em partially} with residual term $\sim \epsilon_{\phi-}$ as well as for amplitude quadratures --- compare with term $\sim \epsilon_{a+}$ \eqref{beta-acomb3} for $\tilde \beta_{a-}$.

In general case one can measure in each output a pair of quadrature components with arbitrary parameter $\varphi$
\begin{subequations}
 \label{quadphi}
\begin{align}
 b_{+\varphi} &= b_{+a}\cos\varphi +b_{+\phi}\sin\varphi,\\
   b_{-\varphi}&= b_{-a}\cos\varphi -b_{-\phi}\sin\varphi
\end{align}
\end{subequations}
It is easy to show that sum $(b_{+\varphi}+ b_{-\varphi})$ is not disturbed by the mechanical motion but contains the term proportional to the back action force, whereas the difference $(b_{+\varphi}- b_{-\varphi})$ contains the term proportional to mechanical motion (with back action and signal). The back action term can be {\em partially} measured and subtracted from the force measurement result.

\section {Conventional (degenerate) squeezing}

Let consider separately the case of conventional {\em (degenerate)} squeezing in each mode $\omega_\pm=\omega_0\pm \omega_m$, for example, see \cite{Walls2008}. For such squeezing parameteric pumps on frequencies $2\omega_\pm$ should be realized. We use the same Hamiltonian \eqref{Halt}, excepting the nonlinear squeezing Hamiltonian  $H_{sq}$, which we write in form 
\begin{subequations}
\begin{align}
    H_{sq} &= \frac{\hslash \nu_-}{i}\left( c_{01}^* c_-^{ 2} -c_{01} c_-^{\dag 2} \right) +\\
    &\qquad +\frac{\hslash \nu_+}{i}\left(
   c_{02}^* c_+^{ 2} - c_{02} c_+^{\dag 2} \right),
\end{align}   
\end{subequations}
where $\nu_\pm$ is a constants, describing nonlinearity  of degenerate parametric amplification in modes,  $c_{01},\ c_{02}$ --- are annihilation operators  of pumps on frequencies $2\omega_-=2(\omega_0-\omega_m)$ and $2\omega_+=2(\omega_0-\omega_m)$ correspondingly. Below we assume equality of pump amplitudes and nonlinearities 
\begin{subequations}
 \label{assume}
\begin{align}
  c_{01} &= c_{02}=C_{00}, \\
  \nu_+ &=\nu_-\equiv \nu, \quad \upsilon = 2\nu C_{00},
\end{align}
\end{subequations}
and introduce notation for $\upsilon$.

Again, the sets for amplitude quadratures and for phase quadratures are independent and can be separated. 
As example, we consider the set for amplitude quadratures. The main difference, as compared with \eqref{quadIn2}, is that {\em both} $g_{a+}$ and $g_{a-}$ are squeezed by similar way, whereas in \eqref{quadIn2} --- by opposite way ($g_{a+}$ is unsqueezed and $g_{a-}$ is squeezed) --- see details of derivation in Appendix \ref{appC}. For sum and difference of output amplitude quadratures we obtain:
\begin{subequations}
 \label{betaUSq}
 \begin{align}
 \beta_{a+} &= \zeta \alpha_{a+} + \sigma \sqrt \frac{\gamma_e}{\gamma_0} \epsilon_{a_+},\\  
 \beta_{a-} &=\frac{\sqrt{\mathcal N}}{(\gamma_m -i\Omega)} \left\{\left(\zeta\alpha_{a-} +\sigma \sqrt\frac{\gamma_e}{\gamma_0}\,\epsilon_{a-} \right) \frac{(\gamma_m -i\Omega)}{\sqrt{\mathcal N}} \right.-\nonumber\\
 \label{betaUSqBA-1}
 & \quad -\sqrt{\mathcal N} \left(\alpha_{a+} +\sqrt\frac{\gamma_e}{\gamma_0}\, \epsilon_{a+}\right) -\\
 \label{betaUSqBA-2}
 &\quad \left. - \Big(\sqrt {2 \gamma_m}\, q_a +  f_{s\,a}\Big)\right\},\\
 & \text{where}\nonumber \\ \label{N0}
&\mathcal N = \frac{\gamma^2 \mathcal N_0}{(\gamma +\upsilon - i\Omega)^2},\quad \mathcal N_0=\frac{4\gamma_0\, \eta^2 C_0^2}{\gamma^2},\\ 
\label{zeta}
&\zeta=\frac{\gamma_0 -\gamma_e -\upsilon+ i\Omega }{\gamma_0 +\gamma_e +\upsilon - i\Omega },\\
&\sigma = \frac{2\gamma_0  }{\gamma_0 +\gamma_e +\upsilon - i\Omega }\,
 \end{align}
\end{subequations}
Using (\ref{betaUSqBA-1}, \ref{betaUSqBA-2}) we derive power spectral density for this case 
\begin{subequations}
\label{SfcC}
\begin{align}
    S_{f}^{\beta_{a-}} (\Omega) &= 2\gamma_m(2 n_T + 1) + \\
    &\qquad+ \frac{\gamma_m^2 + \Omega^2}{|N|} \left( |\zeta|^2 + |\sigma|^2 \frac{\gamma_e}{\gamma_0} \right) + \\ 
    &\qquad+ |N| \left( 1 + \frac{\gamma_e}{\gamma_0} \right)
\end{align}
\end{subequations}
\begin{figure}
    \includegraphics[width=0.45\textwidth]{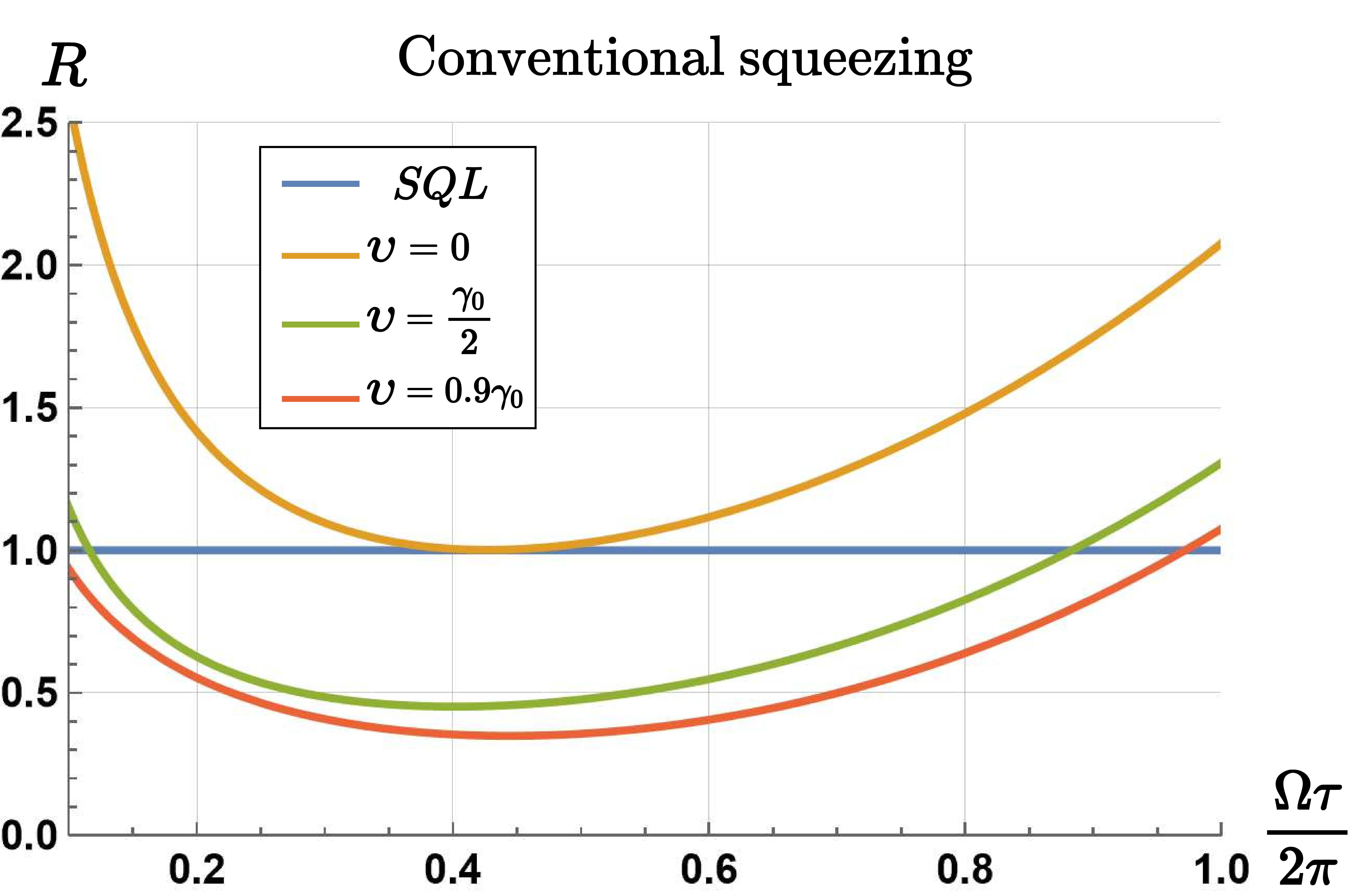}
    \caption{Plots of ratio $R=S_{f}^{\beta_{a-}}/S_{SQL,f}$ of spectral densities (\ref{SfcC}, \ref{SSQL}) as function of spectral frequency $\Omega$ at conditions $\gamma_e \neq 0, \, \upsilon \neq 0$ and $\gamma_m=0$. Dimensionless power $\mathcal N_0= \pi/\tau$ \eqref{N0}, $\tau$ is time of signal force action. Other parameters are taken from Table~\ref{table1}. \label{WithBACS}}
\end{figure}
Comparing the result in Fig.~\ref{WithBACS} with the one obtained earlier for two-photon squeezing Fig.~\ref{WithBA}, we can notice that the curves almost coincide. However, in the case of two-photon squeezing the curve is slightly lower, we assume that this may be due to the correlation of photons in the case of two-photon squeezing.

Two port measurement (detection of $\beta_{a+},\ \beta_{a-}$ separately) allows to subtract back action term in \eqref{betaUSqBA-1} (post processing subtraction), however, due to optical losses it can be done {\em partially only}: 
\begin{subequations}
\label{tildebetaa-}
\begin{align}
   &\left[\alpha_{a+} +\sqrt\frac{\gamma_e}{\gamma_0}\, \epsilon_{a+} \right] \  \Rightarrow\\
   &\Rightarrow\ \left[\alpha_{a+} +\sqrt\frac{\gamma_e}{\gamma_0}\, \epsilon_{a+} \right] - \left(\alpha_{a+} + \frac{\sigma}{\zeta} \sqrt \frac{\gamma_e}{\gamma_0}\, \epsilon_{a_+}\right) = \nonumber\\
   &=-\frac{1}{\zeta} \sqrt \frac{\gamma_e}{\gamma_0}\, \epsilon_{a_+}
 \end{align}
 \end{subequations}
 It means one can measure combination
 \begin{subequations}
 \label{tildebetaUSq}
\begin{align}
   \tilde \beta_{a-} &= \frac{\sqrt{\mathcal N}}{(\gamma_m -i\Omega)} \left\{\left(\zeta\alpha_{a-} +\sigma \sqrt\frac{\gamma_e}{\gamma_0}\,\epsilon_{a-} \right) \frac{(\gamma_m -i\Omega)}{\sqrt{\mathcal N}} \right.-\nonumber\\
\label{tildebetaUSqBA}
 & \quad +\sqrt{\mathcal N} \left(\frac 1 {\zeta} \sqrt\frac{\gamma_e}{\gamma_0}\, \epsilon_{a+}\right)-\\
 &\quad \left.- \Big(\sqrt {2 \gamma_m}\, q_a +  f_{s\,a}\Big)\right\}
 \end{align}
\end{subequations}
Back action term \eqref{tildebetaUSqBA} is depressed as compare with \eqref{betaUSqBA-1}.

Using \eqref{tildebetaUSq} one can derive power spectral density, recalculated to normalized  signal force $f_{s\,a}$ for the case of vacuum input fluctuations $\alpha_{a\pm}, \epsilon_{a\pm}$
\begin{subequations}
 \label{SfUSq}
\begin{align}
   S_{f\,US} &= \left(|\zeta|^2+|\sigma|^2 \frac{\gamma_e}{\gamma_0}\right) \frac{|\gamma_m -i\Omega|^2}{|\mathcal N|} \\
\label{SfUSqBA}
 & \quad + \left(\frac{|\mathcal N|} {|\zeta|^2}\, \frac{\gamma_e}{\gamma_0}\right)+2 \gamma_m\big(2n_T+1\big) 
 \end{align}
\end{subequations}
Here first term describes measurement error, second term --- residual back action and last one corresponds to thermal fluctuations.

\begin{figure}
    \includegraphics[width=0.45\textwidth]{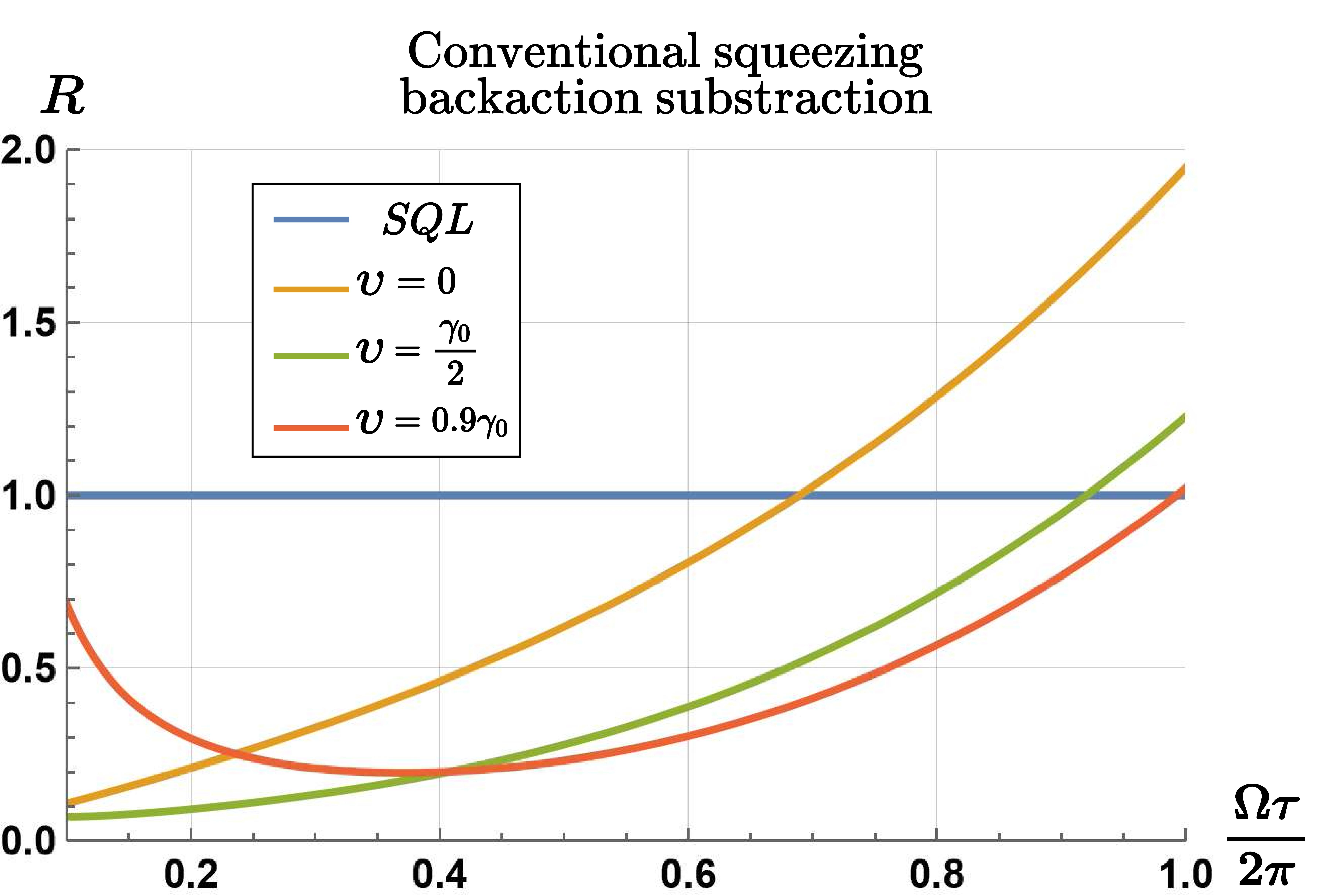}
    \caption{Plots of ratio $R=S_{f\, US}/S_{SQL,f}$ of spectral densities (\ref{SfUSq}, \ref{SSQL}) as function of spectral frequency $\Omega$ at conditions $\gamma_e \neq 0, \, \upsilon \neq 0$ and $\gamma_m=0$. Dimensionless power $\mathcal N_0= \pi/\tau$ \eqref{N0}, $\tau$ is time of signal force action. Other parameters are taken from Table~\ref{table1}. \label{NoBACS}}
\end{figure}
\begin{figure}
    \includegraphics[width=0.45\textwidth]{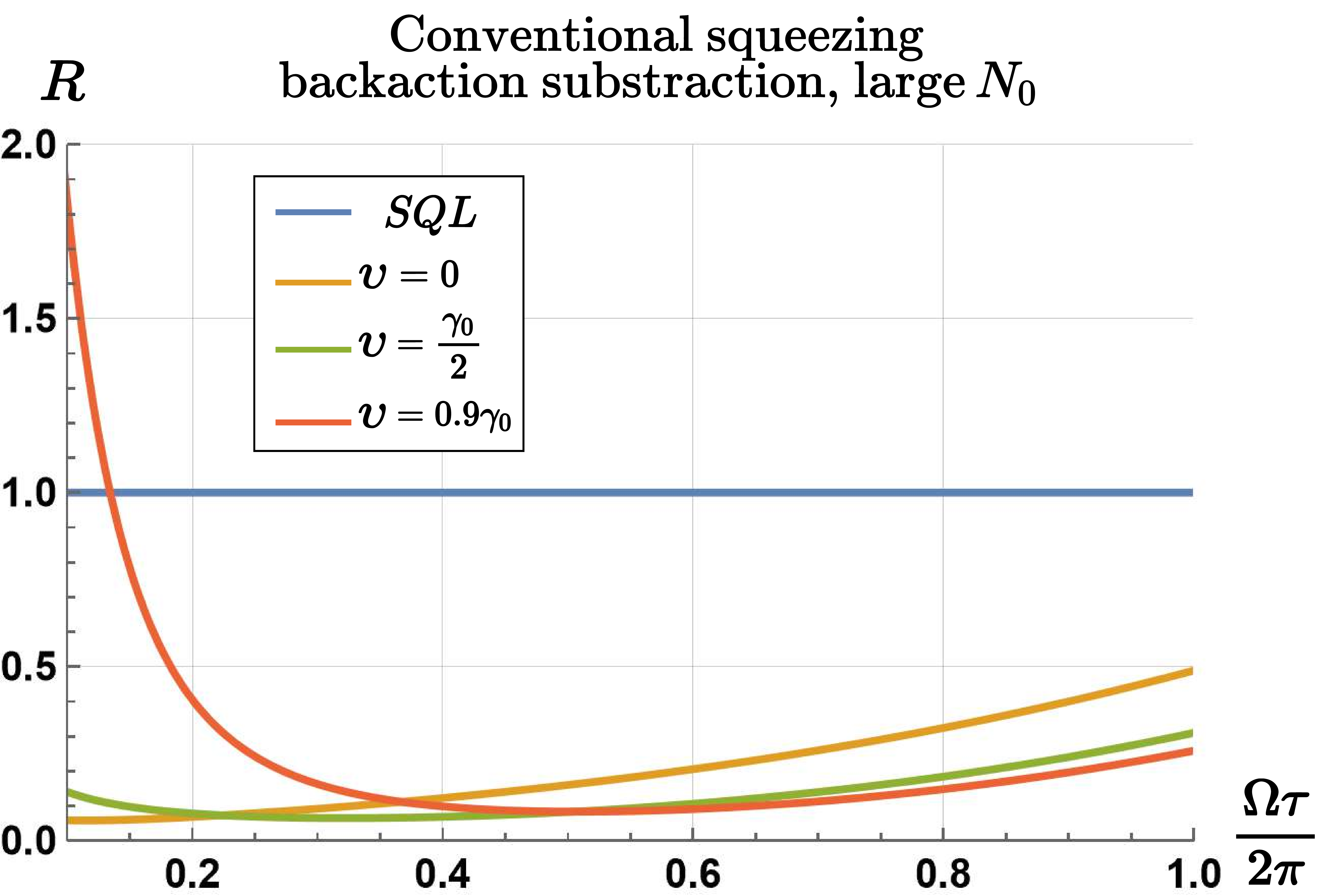}
    \caption{Plots of ratio $R=S_{f\, us}/S_{SQL,f}$ of spectral densities (\ref{SfUSq}, \ref{SSQL}) as function of spectral frequency $\Omega$ at conditions $\gamma_e \neq 0, \, \upsilon \neq 0$ and $\gamma_m=0$. Here, in contrast to Fig.~\ref{WithBACS}, we take a large dimensionless power: $\mathcal N_0= 4\pi/\tau$. Other parameters are the same. \label{NoBACS2}}
\end{figure}

The spectral density with subtraction of the back action \eqref{SfUSq} is shown in Fig.~\ref{NoBACS}.
In contrast to two-photon squeezing Fig.~\ref{NoBA} the subtraction of the back action in the case of one-photon squeezing Fig.~\ref{NoBACS} does not give such a considerable gain at low frequencies. This can be explained by the fact that the optical losses noise, defined by terms $\sim \epsilon_{a\pm}$ (this term is $\sim \mathcal N$ in \eqref{SfUSqBA}), on low frequencies produce {\em larger} contribution for conventional squeezing, than for two-photon squeezing. Indeed, noted term in \eqref{SfUSqBA} in inverse proportional to $|\zeta|^2$  which became small at small frequency $\Omega\to 0$ and large squeezing factor $\upsilon \to \gamma_0+\gamma_e$, see \eqref{zeta}. In contrast, in spectral density for two photon nondegenerate squeezing \eqref{Sfd} analogous term \eqref{SfbaRes} is proportional to $|\xi_-/\xi_+^2|$, which does not increase in case $\Omega\to 0,\ \kappa\to \gamma_0+\gamma_e$, see definition \eqref{xipm}. It is evidence of more stability of two photon squeezing to optical losses as compared with conventional squeezing. For illustration on Fig.~\ref{NoBACS2} we present plots for 4 times larger pump $\mathcal N_0=4\pi/\tau$ as compared with plots on Fig.~\ref{NoBACS} --- for large squeezing factor $\upsilon=0.9\,\gamma_0$ spectral density dramatically increases at low $\Omega$.

\section{Discussion and Conclusion}

We have investigated a broadband multidimensional variation measurement, proposed in \cite{22PRAVyNaMa}, with account of optical losses and two kinds of intracavity squeezing: two photon (nondegenerate) and conventional (degenerate) one. 


 We show that two photon nondegenerate internal squeezing, realized in two modes of cavity, demonstrates that, for example, of sum of amplitudes quadratures is squeezed whereas its difference is unsqueezed (or vice versa). (The same is valid for measurement of phase quadratures.) This squeezing can be registered in complete form only in ``two port'' detection when output amplitude quadratures of {\em each} mode are detected {\em separately} by balanced homodyne detectors with local oscillators frequencies $\omega_\pm$. Recall, usually the detection of two photon squeezing is realized with one local oscillator wave (see, for example, Sec. 5.2 in \cite{Walls2008}). ``Two port'' detection is formally similar to situation in broadband variational measurement. 

We do not consider how the {\em external} degenerate and nondegenerate squeezing can improve the sensitivity. However, we plan to analyze external squeezing in future because of obtained formulas can be easy applied fot it.

Account of optical losses gives possibility to more real estimates for experimental realization. In particular,  back action subtraction is not complete in presence of optical losses, noise due to losses noise restricting value of subtraction.

We also demonstrate that conventional (degenerate) squeezing in each mode $\omega_\pm$ and two photon (nondegenerate) squeezing in case of zero  optical losses give similar results. However in presence of optical losses  the value of back action subtraction  for conventional squeezing is worse than for two photon squeezing.


%
\begin{figure}
 \includegraphics[width=0.45\textwidth]{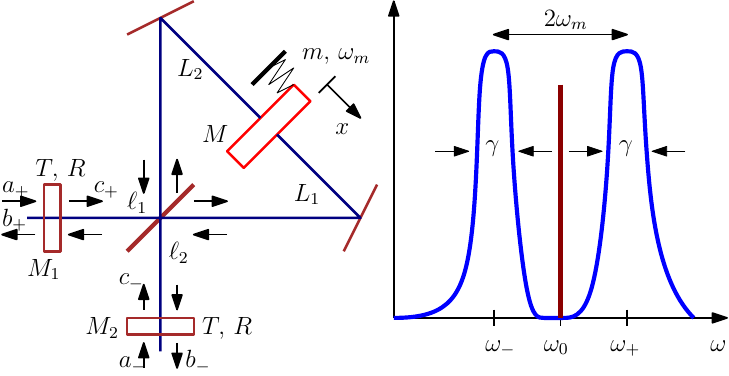}
 \caption{Schematic of the Michelson-Sagnac interferometer in which mirror $M$ is completely reflecting. The mirror is a test mass $m$ of the mechanical oscillator with frequency $\omega_m$. Two eigenmodes with frequencies $\omega_-,\ \omega_+$ are coupled with the mechanical oscillator. Relaxation rate $\gamma$ is the same for the both modes, $\gamma\ll \omega_m$. The frequency of non-resonant pump is $\omega_0=(\omega_+ + \omega_-)/2$. }\label{scheme22}
\end{figure}

As a possibility of experimental realization one can use {\em two} modes scheme with frequencies $\omega_0\pm \omega_m$ and usage of the pump with frequency $\omega_0$ located in between of the modes --- see Fig.~\ref{scheme22}. The system  has two degenerate modes. If the position of a perfectly reflecting mirror $M$ is fixed, one MSI mode, characterized with frequency $\omega_+$, is given by a light wave which travels between $M_1$ and BS. The light is split on the BS and after reflection from mirror $M$ returns exactly to $M_1$. It does not propagate to $M_2$. For the other mode, characterized with frequency $\omega_-$,  wave  travels from $M_2$ to BS and after reflection from $M$ returns to $M_2$ and does not propagate to $M_1$. The frequencies of the modes, $\omega_\pm$, are controlled by variation of path distances $\ell_1,\ \ell_2$. Small shift of mirror $M$ position provides coupling between the modes.  Mirror $M$ is a test mass of the mechanical oscillator with mass $m$ and eigenfrequency $\omega_m$. The back action can be suppressed in this scheme as well. However, pump on frequency $\omega_0$ is not resonant and more optical power will be needed to beat the SQL. In addition pump should be excited though both mirrors $M_1,\ M_2$.

We hope that analyzed here two photon nondegenerate squeezing in broadband coherent multidimensional variational measurement can be used in precision optomechanical measurements including laser gravitational wave detectors.

 \acknowledgments
The research of SPV has been supported by  by Theoretical Physics and Mathematics Advancement Foundation “BASIS” (Contract No. 22-1-1-47-1), by the Interdisciplinary Scientific and Educational School of Moscow University ``Fundamental and Applied Space Research'' and by the TAPIR GIFT MSU Support of the California Institute of Technology.  This document has LIGO number P2300325.   

\appendix

\section{Description of relaxation}
\label{IntrDeriv}
 
In this appendix we provide details of the standard calculation for intracavity fields, for example, see \cite{Walls2008}.

We begin with Hamiltonian \eqref{Halt}
\begin{align}  
   \label{Ht}
      H_{T,\,0} &= \sum\limits_{k=0}^\infty \hslash \omega_k a_k^\dag a_k,\\
      H_{\gamma_0} & =i \hslash \sqrt\frac{\gamma_0 \Delta \omega}{\pi} \sum\limits_{k=0}^\infty \left[(c_+^\dag+c_-^\dag) a_k-(c_++c_-)a_k^\dag \right],\nonumber\\
       H_{T,\,e} &= \sum\limits_{k=0}^\infty \hslash \omega_k e_k^\dag e_k,\\
      H_{\gamma_e} & =i \hslash \sqrt\frac{\gamma_e \Delta \omega}{\pi} \sum\limits_{k=0}^\infty \left[(c_+^\dag+c_-^\dag) e_k-(c_++c_-)e_k^\dag \right],\nonumber\\
       \label{Htm}
      &H_{T, \, m} = \sum\limits_{k=0}^\infty \hslash \omega_k q_k^\dag q_k,\\
      \label{Hgammam}
      &H_{\gamma_m}  =i \hslash \sqrt\frac{\gamma_m \Delta \omega}{\pi} \sum\limits_{k=0}^\infty \left[d^\dag q_k-dq_k^\dag \right].
  \end{align}
Here $H_{T,\,0}$ is the Hamiltonian of the environment presented as a bath of oscillators \footnote{Here and below we consider the equidistant spectrum for all thermal baths.} described with frequencies $\omega_k=\omega_{k-1}+\Delta \omega$ and annihilation and creation operators $a_k$, $a_k^\dag$.  $H_{\gamma_0}$ is the Hamiltonian of coupling between the environment and the optical resonator, $\gamma_0$ is the coupling constant. Physical sense of $a_k$ are amplitudes of modes falling on input mirror of cavity. $H_{T,\,e}$ and $H_{\gamma_e}$ are the Hamiltonians described another environment and interaction with resonator corresponding to optical losses.    Similarly $H_{T, \, m}$ is the Hamiltonian of the environment presented by a thermal bath of mechanical oscillators with frequencies $\omega_k=\omega_{k-1}+\Delta \omega$ and amplitudes described with annihilation and creation operators $q_k$, $q_k^\dag$.  $H_{\gamma_m}$ is the Hamiltonian of coupling between the environment and the mechanical oscillator, $2\gamma_m$ is the decay rate of the oscillator.
  
As a first step we account only $H_{T,\,0},\ H_{\gamma_0}$. Heisenberg equations for operators $c_+$ and $a_k$ are the following:
\begin{subequations} 
  \begin{align}
 \dot c_+ & =-i\omega_+ c_+- \eta c_0 d+ \sqrt\frac{\gamma_0 \Delta \omega}{\pi} \sum\limits_{k=0}^\infty a_k,\\
\dot a_k & =-i \omega_k a_k-\sqrt\frac{\gamma_0 \Delta \omega}{\pi}  \left(c_++c_-\right)
  \end{align} \label{Heis1}
  \end{subequations}

We introduce slow amplitudes $c_\pm \rightarrow c_\pm e^{-i \omega_\pm t}$, $d \rightarrow d e^{-i\omega_m t}$, $a_k \rightarrow a_k e^{-i \omega_k t}$ and substitute them into \eqref{Heis1}
\begin{subequations} 
  \begin{align}
\dot c_+ & =-\eta c_0 d+  \sqrt\frac{\gamma_0 \Delta \omega}{\pi} \sum\limits_{k=0}^\infty a_k e^{-i(\omega_k-\omega_+)t}, \label{Heisc}\\
  \dot a_k & = -\sqrt\frac{\gamma_0 \Delta \omega}{\pi} \left(c_+e^{-i\omega_+t}+c_- e^{-i\omega_-t}\right)e^{i\omega_kt} \label{Heisb}
  \end{align}
  \end{subequations}

Using initial condition $a_k(t=0)=a_k(0)$ to integrate \eqref{Heisb} we derive
\begin{align}
a_k(t) & =a_k(0)-\int\limits_0^t\sqrt\frac{\gamma_0 \Delta \omega}{\pi}  c_+(s)e^{-i(\omega_+-\omega_k)s}ds-\\
& -\int\limits_0^t\sqrt\frac{\gamma_0 \Delta \omega}{\pi}c_-(s)e^{-i(\omega_--\omega_k)s}ds \label{binit}
\end{align} 
Using the condition $a_k(t=\infty)=a_k(\infty)$ to integrate \eqref{Heisb} we derive
\begin{align}
a_k(t) & =a_k(\infty)+\int\limits_t^\infty\sqrt\frac{\gamma_0 \Delta \omega}{\pi}  c_+(s)e^{-i(\omega_+-\omega_k)s}ds+\\
& +\int\limits_t^\infty\sqrt\frac{\gamma_0 \Delta \omega}{\pi}c_-(s)e^{-i(\omega_--\omega_k)s}ds \label{bfin}
\end{align} 
To get the input-output relation we substitute initial condition \eqref{binit} into \eqref{Heisc}
\begin{subequations} 
  \begin{align}
\dot c_+ & =-\eta c_0 d+   \sum\limits_{k=0}^\infty \sqrt\frac{\gamma_0 \Delta \omega}{\pi} a_k(0)e^{-i(\omega_k-\omega_+)t}-\\
&-\sum\limits_{k=0}^\infty\int\limits_0^t\frac{\gamma_0 \Delta \omega}{\pi}  c_+(s) 
	 e^{-i(\omega_k-\omega_+)(t-s)}ds -\\
&-\left(\sum\limits_{k=0}^\infty\int\limits_0^t\frac{\gamma_0 \Delta \omega}{\pi}  c_-(s) e^{-i(\omega_k-\omega_-)(t-s)}ds \right)e^{i(\omega_+-\omega_-)t}\nonumber
  \end{align} \label{Heis2}
  \end{subequations}
and omit the last term proportional to $e^{i(\omega_+-\omega_-)t}$ as the fast oscillating, and define the input field
  \begin{equation}
a_+(t) = \sum\limits_{k=0}^\infty \sqrt\frac{ \Delta \omega}{2\pi} a_k(0)e^{-i(\omega_k-\omega_+)t} \label{aplus}
\end{equation}
To calculate the remaining sum in \eqref{Heis2} we  assume the validity of limit $\Delta \omega \rightarrow 0$ and  replace the sum by the integration
\begin{subequations} 
  \begin{align} 
&\sum\limits_{k=0}^\infty\int\limits_0^t\frac{\gamma_0 \Delta \omega}{\pi}  c_+(s) e^{-i(\omega_k-\omega_+)(t-s)}ds  \rightarrow\\ \nonumber
\rightarrow & \int\limits_0^\infty\int\limits_0^t 2\gamma_0  c_+(s) e^{-i(\omega-\omega_+)(t-s)}ds \frac{d \omega}{2 \pi}=\\ \nonumber
= &  \int\limits_{-\omega_+}^\infty\int\limits_0^t 2\gamma_0  c_+(s) e^{-i \omega(t-s)}ds \frac{d \omega}{2 \pi}\approx\\ \nonumber
\approx &  \int\limits_{-\infty}^\infty\int\limits_0^t 2\gamma_0  c_+(s) e^{-i \omega(t-s)}ds \frac{d \omega}{2 \pi}=\\
= & \int\limits_0^t 2\gamma_0  c_+(s) \delta(t-s) ds = \frac{2 \gamma_0 c_+(t)}{2}=\gamma_0 c_+(t)
  \end{align} \label{gammac}
  \end{subequations}

Substituting \eqref{aplus} and \eqref{gammac} into \eqref{Heis2} we obtain
\begin{align}
\dot c_+ & =-\eta c_0 d+  \sqrt{2\gamma_0}a_+-\gamma_0 c_+. \label{cplus}
\end{align}
Taking into account Hamiltonians $H_{T,\, e},\ H_{\gamma_e}$ by similar way we obtain generalization of \eqref{cplus}
\begin{align}
	\dot c_+ & =-\eta c_0 d+  \sqrt{2\gamma_0}a_+ +\sqrt{2\gamma_e} e_+-(\gamma_0+\gamma_e) c_+. \label{cplus2}
\end{align}

By analogue, we derive the equation for input field $a_-$ and present it in a similar form
  \begin{equation}
a_-(t) = \sum\limits_{k=0}^\infty \sqrt\frac{ \Delta \omega}{2\pi} a_k(0)e^{-i(\omega_k-\omega_-)t}
\end{equation}
It leads to the equation for the intracavity field $c_-$
   \begin{align}
\dot c_-+ \gamma_0 c_--\eta c_0 d^\dag=  \sqrt{2\gamma_0}a_-. \label{cminus}
\end{align}
with generalization with account $H_{T,\, e},\ H_{\gamma_e}$: \eqref{cplus}
\begin{align}
	\dot c_- & =-\eta c_0 d^\dag+  \sqrt{2\gamma_0}a_- +\sqrt{2\gamma_e} e_--(\gamma_0+\gamma_e) c_-. \label{cminus22}
\end{align}

Similar equation can be derived for the amplitude $q(t)$ of the mechanical oscillator
  \begin{equation}
q(t)=  \sum\limits_{k=0}^\infty \sqrt\frac{ \Delta \omega}{2\pi} b_{m, \,k}(0)e^{-i(\omega_k-\omega_+)t},
\end{equation}
resulting in the Langevin  equation for mechanical oscillator quadrature $d$
   \begin{align}
\dot d+ \gamma_m d- \eta^* [c_0 c_-^\dag + c_+ c_0^\dag]=  \sqrt{2\gamma_m}q.
\end{align}
To derive the input-output relation we substitute \eqref{bfin} into \eqref{Heisc} and define the output amplitudes 
  \begin{align}
b_+(t) = -\sum\limits_{k=0}^\infty \sqrt\frac{ \Delta \omega}{2\pi} a_k(\infty)e^{-i(\omega_k-\omega_+)t} \\
b_-(t) = -\sum\limits_{k=0}^\infty \sqrt\frac{ \Delta \omega}{2\pi} a_k(\infty)e^{-i(\omega_k-\omega_+)t}.
\end{align}
It leads to 
   \begin{align} 
\dot c_+ & - (\gamma_0+\gamma_e) c_++\eta c_0 d=  \sqrt{2\gamma_0}b_+ \label{bplus}\\
\dot c_- & - (\gamma_0+\gamma_e) c_-+\eta^* c_0 d^\dag=  \sqrt{2\gamma_0}b_- \label{bminus}
\end{align}
Utilizing pairs of equations \eqref{cplus} and \eqref{bplus}  as well as \eqref{cminus} and \eqref{bminus} we obtain the final expression for the input-output relations
\begin{align}
b_+=-a_++\sqrt{2\gamma_0} c_+\\
b_-=-a_-+\sqrt{2\gamma_0} c_-
\end{align}

Let us to derive the commutation relations for the Fourier amplitudes of the operators. We introduce Fourier transform of field $a_+(t)$ using \eqref{aplus}
 \begin{subequations}
 \begin{align}
a_+(\Omega) = \int\limits_{-\infty}^{\infty} \sum\limits_{k=0}^\infty \sqrt\frac{ \Delta \omega}{2\pi} a_k(0)e^{-i(\omega_k-\omega_+-\Omega)t}dt=\\
= \sum\limits_{k=0}^\infty \sqrt { 2 \pi \Delta \omega}  a_k(0)  \delta(\Omega-\omega_k+\omega_+)
 \end{align}
 \end{subequations}
 This allows us to find the commutators \eqref{comm1}: 
  \begin{subequations}
 \begin{align} \nonumber
 [a_+(\Omega)& ,a_+^\dag(\Omega') ] 
   =\sum\limits_{k=0}^\infty  2 \pi \Delta \omega   [a_k(0),a_k^\dag(0)] \times\\ \nonumber
&\times \delta(\Omega-\omega_k+\omega_+)  \delta(\Omega'-\omega_k+\omega_+) \rightarrow\\ \nonumber
&\rightarrow \int\limits_{-\infty}^\infty 2 \pi [a_+(0),a_+^\dag(0)]  \delta(\Omega-\omega)  \delta(\Omega'-\omega) d\omega=\\
&=2\pi \delta(\Omega-\Omega'),
 \end{align}
 \end{subequations}
 and the correlators \eqref{corr1e} (we assume that oscillators of environment are in thermal state, for optics --- in vacuum state)
   \begin{subequations}
 \begin{align} \nonumber
\langle a_+(\Omega)&,a_+^\dag(\Omega') \rangle
 = \sum\limits_{k=0}^\infty  2 \pi \Delta \omega   \langle a_k(0),a_k^\dag(0) \rangle \times\\ \nonumber
&\times \delta(\Omega-\omega_k+\omega_+)  \delta(\Omega'-\omega_k+\omega_+) \rightarrow\\ \nonumber
&\rightarrow \int\limits_{-\infty}^\infty 2 \pi \langle a_+(0)a_+^\dag(0) \rangle \delta(\Omega-\omega)  \delta(\Omega'-\omega) d\omega=\\
&=2\pi \delta(\Omega-\Omega')
 \end{align}
 \end{subequations}
Similar expressions can be derived for commutators and correlators of the optical $a_-$ and mechanical $q$ quantum amplitudes.

\section{Input-output relations (two photon squeezing)}\label{appB}

Here we present detail formulas for derivation of output quadratures \eqref{betaMSIa} for case of two photon squeezing.

 Using definition \eqref{apmFT} and \eqref{commT} we derive commutators and correlators for the Fourier amplitudes of the loss fluctuation operators $e_\pm$ and thermal noise operators $\hat q$
\begin{align}
	\label{comm1e}
	\left[ e_\pm(\Omega),  e_\pm^\dag(\Omega')\right] &= 2\pi\,\delta(\Omega -\Omega'),\\
	\label{corr1e}
	\left\langle e_\pm(\Omega)\,  e_\pm^\dag(\Omega')\right\rangle &= 2\pi\, \delta(\Omega -\Omega'),\\
	\left[ q(\Omega),\,  q^\dag(\Omega')\right] &= 2\pi\,  \delta(\Omega-\Omega'),\\
	\label{corrq}
	\left\langle q(\Omega)\,  q^\dag(\Omega')\right\rangle 
	&= 2\pi\,  \big(2n_T+1\big)\, \delta(\Omega-\Omega').
\end{align}
For operator ($a_\pm(\Omega)$) of input fluctuations formulas are similar.

Using \eqref{set1} one can derive set for the Fourier amplitudes $c_\pm$ for the intracavity fields  as well as mechanical amplitude $d$:
\begin{subequations}
	\label{set2}
	\begin{align}
		(\gamma -i\Omega)c_+(\Omega)&=-\eta C_0  d(\Omega) + \kappa c_-^\dag(-\Omega) +\\
		& \qquad+\sqrt{2 \gamma_0}\, a_+ (\Omega) +\sqrt{2\gamma_e}  e_+(\Omega), \nonumber\\
		(\gamma -i\Omega) c_-(\Omega)&=\eta C_0  d^\dag(-\Omega) + \kappa c_+^\dag(-\Omega) +\\
		&\qquad + \sqrt{2 \gamma_0}\, a_-(\Omega)+\sqrt{2 \gamma_e}\, e_-(\Omega),\nonumber\\
		(\gamma_m -i\Omega) d(\Omega)&=\eta C_0\big[  c_-^\dag(-\Omega) +   c_+(\Omega)\big]+\\
		&\qquad  +\sqrt{2 \gamma_m}\,\hat q(\Omega) +f_s(\Omega).\nonumber
	\end{align}
\end{subequations}

Let introduce quadratures of amplitude and phase \eqref{quadDef}.
Then using \eqref{set2} we obtain
\begin{subequations}
	\label{quadIn}
	\begin{align}
		\label{ca+}   
		(\gamma - i\Omega)c_{+a} -\kappa c_{-a}+  \eta C_0  d_a &=\\
		= \sqrt {2 \gamma_0} a_{ +a}  &+\sqrt {2 \gamma_e} e_{ +a},\nonumber\\
		\label{cphi+}   
		(\gamma  - i\Omega)c_{+\phi} +\kappa c_{-\phi}+  \eta C_0  d_\phi &=\\
		+ \sqrt {2 \gamma_0} a_{ +\phi} &+ \sqrt {2 \gamma_e} e_{ +\phi},\nonumber\\
		\label{ca-} 
		(\gamma - i\Omega) c_{-a} -\kappa c_{+a}- \eta C_0  d_a &= \\
		=\sqrt {2 \gamma_0} a_{-a} &+\sqrt{2\gamma_e} e_{-a},\nonumber\\
		\label{cphi-} 
		(\gamma - i\Omega) c_{-\phi} +\kappa c_{+\phi}+ \eta  C_0  d_\phi &= \\
		=\sqrt {2 \gamma_0} a_{-\phi} &+\sqrt{2\gamma_e} e_{-\phi},\nonumber \\
		\label{da}
		(\gamma_m - i\Omega)  d_a - \eta C_0 \Big(c_{+a}+c_{-a}\Big)&= \\
		= \sqrt {2 \gamma_m} q_a +  f_{s\,a},\nonumber\\
		\label{dphi}
		(\gamma_m - i\Omega)  d_\phi  - \eta C_0  \Big(c_{+\phi} - c_{-\phi}\Big) &=\\
		= \sqrt {2 \gamma_m} q_{\phi} +  f_{s\,\phi}.\nonumber
	\end{align}
\end{subequations}
Please note that sum $(c_{+a} +c_{-a})$ does not contain  information on the mechanical motion (term proportional to $\sim d_a$ is absent), but produces the back action term in \eqref{da}. At the same time difference $(c_{+\phi} -c_{-\phi})$ does not contain information on mechanical motion (term $\sim d_\phi$), but is responsible on back action in \eqref{dphi}. So it should be useful to introduce sum and difference of the quadratures \eqref{gDef} and rewrite \eqref{quadIn} in the new notations
\begin{subequations}
	\label{quadIn2}
	\begin{align}
		\label{ga+}   
		(\gamma -\kappa - i\Omega)g_{a+} = \sqrt {2 \gamma_0} & \alpha_{ a+}
		+\sqrt{2\gamma_e}\epsilon_{a+},\\
		\label{ga-} 
		(\gamma +\kappa- i\Omega) g_{a-} + \sqrt 2 \eta C_0  d_a &= \sqrt {2 \gamma_0}  \alpha_{a-}+\\ &+\sqrt{2\gamma_e}\epsilon_{a-},\nonumber \\
		\label{da2}
		(\gamma_m - i\Omega)  d_a - \sqrt 2 \eta C_0 g_{a+} 
		&= \sqrt {2 \gamma_m} q_a +  f_{s\,a},\\
		\label{gphi+}   
		(\gamma +\kappa - i\Omega)g_{\phi+} +  \sqrt 2\eta C_0  d_\phi &= \sqrt {2 \gamma_0}  \alpha_{ \phi+}+\\
		& +\sqrt{2\gamma_e}\epsilon_{\phi+},\nonumber\\
		\label{gphi-} 
		(\gamma -\kappa - i\Omega) g_{\phi-}  = \sqrt {2 \gamma_0} & \alpha_{\phi-}
		+\sqrt{2\gamma_e}\epsilon_{\phi-},\\
		\label{dphi2}
		(\gamma_m - i\Omega)  d_\phi  - \sqrt 2 \eta C_0g_{\phi-} 
		&= \sqrt {2 \gamma_m} q_{\phi} +  f_{s\,\phi}.
	\end{align}
\end{subequations}
Obviously, the sets (\ref{ga+}, \ref{ga-}, \ref{da2}) for amplitude quadratures and (\ref{gphi+}, \ref{gphi-}, \ref{dphi2}) for phase quadratures are independent and can be separated.

Using \eqref{outputT} and \eqref{gDef} one can derive equations \eqref{betaMSIa} for sum and difference amplitude quadratures $\beta_{a\pm}$.

\section{Standard Quantum Limit}\label{appSQL}

Here we discuss formula \eqref{SSQL} and present details on derivation of SQL for force acting on mechanical oscillator.

We start from single-sided power spectral density \eqref{Sf} of signal force quadrature $f_{s\, a}$ of normalized signal force  (\ref{Fs}, \ref{fs}), acting during time $\tau$. Approximate condition of detection is 
\begin{align}
 \label{approximate}
    f_{s\, a}&=\frac{F_{s0}}{\sqrt 2\sqrt{2\hslash m\omega_m}} \ge\sqrt{\int_0^{2\pi/\tau} S_f(\Omega)\,\frac{d\Omega}{2\pi} }=\\
    =&\sqrt{\left[2\gamma_m(2n_T+1) +\frac{\gamma_m^2 + \frac{1}{3}\left[\frac{2\pi}{\tau}\right]^2}{\mathcal K} + \mathcal K\right] \frac{1}{\tau}} 
\end{align}
Here we assume that angle $\psi_f=0$ in \eqref{Fs}, and $\mathcal K$ is a constant. We consider the  case of short time $\tau$ of force action: \begin{equation}
 \label{condtau}
 \gamma_m\tau\ll 1   
\end{equation}
 (opposite case of large $\tau$ is not interesting, thermal limit restricts sensitivity). 

Taking minimum over pump $\mathcal K$ with account of \eqref{condtau}, we obtain minimum force $F_{s0}$ to be detected:
\begin{align}
 \label{Fs0}
 \frac{F_{s0}^2}{4\hslash m \omega_m}&\ge  2\gamma_m(2n_T+1)\,\frac{1}{\tau} +\frac{2}{\sqrt 3}\frac{2\pi}{\tau^2}
\end{align}
Here first term describes thermal limit whereas second one --- SQL:
\begin{align}
    F_{s0}^{SQL} =\frac{4}{\tau}\sqrt\frac{\pi \hslash m \omega_m}{\sqrt 3}
\end{align}
This formula is valid with accuracy of constant multiplier about $1$ due to approximation of \eqref{approximate}.

We can take \eqref{SSQL} instead of \eqref{SQu}, then we obtain condition \eqref{condtau}
\begin{align}
 \label{Fs02}
 \frac{\tilde F_{s0}^2}{4\hslash m \omega_m}&\ge  2\gamma_m(2n_T+1)\,\frac{1}{\tau} +\frac{4\pi}{\tau^2}
\end{align}
Last terms in \eqref{Fs0} and in \eqref{Fs02} differ only by multiplier about $1$. So in frequency domain we can use spectral density \eqref{SSQL} for SQL characterization. 

\section{Input-output relations (conventional squeezing)}\label{appC}

Here we present detail formulas for derivation of output quadratures \eqref{betaUSq} for case of standard squeezing.

We obtain set for intracavity amplitudes in frequency domain:
\begin{subequations}
	\label{set2USq}
	\begin{align}
		(\gamma -i\Omega)c_+(\Omega)&=-\eta C_0  d(\Omega) -\upsilon c_+^\dag(-\Omega) +\\
		& \qquad+\sqrt{2 \gamma_0}\, a_+ (\Omega) +\sqrt{2\gamma_e}  e_+(\Omega), \nonumber\\
		(\gamma -i\Omega) c_-(\Omega)&=\eta C_0  d^\dag(-\Omega) -\upsilon c_-^\dag(-\Omega) +\\
		&\qquad + \sqrt{2 \gamma_0}\, a_-(\Omega)+\sqrt{2 \gamma_e}\, e_-(\Omega),\nonumber\\
		(\gamma_m -i\Omega) d(\Omega)&=\eta C_0\big[  c_-^\dag(-\Omega) +   c_+(\Omega)\big]+\\
		&\qquad  +\sqrt{2 \gamma_m}\,\hat q(\Omega) +f_s(\Omega).\nonumber
	\end{align}
 These equations differ from set \eqref{set2} only by terms $\sim \upsilon$.
\end{subequations}

Then using \eqref{set2USq} we obtain for sum and difference of quadratures
\begin{subequations}
	\label{quadInUSq}
	\begin{align}
		\label{ca+USq}   
		(\gamma - i\Omega)c_{+a} +\upsilon c_{+a}+  \eta C_0  d_a &=\\
		= \sqrt {2 \gamma_0} a_{ +a}  &+\sqrt {2 \gamma_e} e_{ +a},\nonumber\\
		\label{cphi+USq}   
		(\gamma  - i\Omega)c_{+\phi} -\upsilon c_{+\phi}+  \eta C_0  d_\phi &=\\
		+ \sqrt {2 \gamma_0} a_{ +\phi} &+ \sqrt {2 \gamma_e} e_{ +\phi},\nonumber\\
		\label{ca-USq} 
		(\gamma - i\Omega) c_{-a} +\upsilon c_{-a}- \eta C_0  d_a &= \\
		=\sqrt {2 \gamma_0} a_{-a} &+\sqrt{2\gamma_e} e_{-a},\nonumber\\
		\label{cphi-USq} 
		(\gamma - i\Omega) c_{-\phi} -\upsilon c_{-\phi}+ \eta  C_0  d_\phi &= \\
		=\sqrt {2 \gamma_0} a_{-\phi} &+\sqrt{2\gamma_e} e_{-\phi},\nonumber \\
		\label{da2c}
		(\gamma_m - i\Omega)  d_a - \eta C_0 \Big(c_{+a}+c_{-a}\Big)&= \\
		= \sqrt {2 \gamma_m} q_a +  f_{s\,a},\nonumber\\
		\label{dphi2c}
		(\gamma_m - i\Omega)  d_\phi  - \eta C_0  \Big(c_{+\phi} - c_{-\phi}\Big) &=\\
		= \sqrt {2 \gamma_m} q_{\phi} +  f_{s\,\phi}.\nonumber
	\end{align}
\end{subequations}
This set differs from \eqref{quadIn} only by terms $\sim \upsilon$. 

Again, sum $(c_{+a} +c_{-a})$ does not contain  information on the mechanical motion (term proportional to $\sim d_a$ is absent), but produces the back action term in \eqref{da2c}. At the same time difference $(c_{+\phi} -c_{-\phi})$ does not contain information on mechanical motion (term $\sim d_\phi$), but is responsible on back action in \eqref{dphi2c}. So it should be useful to introduce sum and difference of the quadratures \eqref{SumDif} and rewrite \eqref{quadInUSq} in the new notations
\begin{subequations}
	\label{quadInUSq2}
	\begin{align}
		\label{ga+2}   
		(\gamma +\upsilon - i\Omega)g_{a+} = \sqrt {2 \gamma_0} & \alpha_{ a+}
		+\sqrt{2\gamma_e}\epsilon_{a+},\\
		\label{ga-2} 
		(\gamma +\upsilon- i\Omega) g_{a-} + \sqrt 2 \eta C_0  d_a &= \sqrt {2 \gamma_0}  \alpha_{a-}+\\ &+\sqrt{2\gamma_e}\epsilon_{a-},\nonumber \\
		\label{da2b}
		(\gamma_m - i\Omega)  d_a - \sqrt 2 \eta C_0 g_{a+} 
		&= \sqrt {2 \gamma_m} q_a +  f_{s\,a},\\
		\label{gphi+2}   
		(\gamma -\upsilon - i\Omega)g_{\phi+} +  \sqrt 2\eta C_0  d_\phi &= \sqrt {2 \gamma_0}  \alpha_{ \phi+}+\\
		& +\sqrt{2\gamma_e}\epsilon_{\phi+},\nonumber\\
		\label{gphi-2} 
		(\gamma -\upsilon - i\Omega) g_{\phi-}  = \sqrt {2 \gamma_0} & \alpha_{\phi-}
		+\sqrt{2\gamma_e}\epsilon_{\phi-},\\
		\label{dphi2b}
		(\gamma_m - i\Omega)  d_\phi  - \sqrt 2 \eta C_0g_{\phi-} 
		&= \sqrt {2 \gamma_m} q_{\phi} +  f_{s\,\phi}.
	\end{align}
\end{subequations}
Again, the sets (\ref{ga+2}, \ref{ga-2}, \ref{da2b}) for amplitude quadratures and (\ref{gphi+2}, \ref{gphi-2}, \ref{dphi2b}) for phase quadratures are independent and can be separated.

As example, let consider the set for amplitude quadratures. The main difference, as compared with \eqref{quadIn2}, is that {\em both} $g_{a+}$ and $g_{a-}$ are squeezed by similar way, whereas in \eqref{quadIn2} --- by opposite way ($g_{a+}$ is unsqueezed and $g_{a-}$ is squeezed).

Using \eqref{outputT} and \eqref{gDef} one can derive equation for $\beta_{a-}$
\begin{subequations}
 \label{betaUSq2}
 \begin{align}
  \beta_{a\pm} &=\sqrt{2\gamma_0}\, \alpha_{a\pm} - \alpha_{a\pm},\quad \gamma\equiv \gamma_0+\gamma_e \\
 \label{diff_betaa+}
  \beta_{a+} &= \frac{\gamma_0 -\gamma_e - \upsilon+ i\Omega }{\gamma_0 +\gamma_e +\upsilon - i\Omega }\cdot \alpha_{ a+} +  \frac{2\sqrt{\gamma_0\gamma_e}\, \epsilon_{ a+}}{\gamma +\upsilon - i\Omega},\\ 
  \label{diff_betaa-}
  \beta_{a-} &=\frac{\gamma_0 -\gamma_e -\upsilon+ i\Omega }{\gamma_0 +\gamma_e +\upsilon - i\Omega }\cdot \alpha_{ a-} +  \frac{2\sqrt{\gamma_0\gamma_e}\, \epsilon_{a-}}{\gamma +\upsilon - i\Omega}-\\
  &\qquad -\frac{2\sqrt \gamma_0\,\eta C_0}{(\gamma +\upsilon - i\Omega)} \,  d_a \nonumber
  \end{align}
  \end{subequations}

Substituting \eqref{da2b} (with \eqref{diff_betaa+}) into \eqref{diff_betaa-}, we finally rewrite equation for $\beta_{a-}$ 
\begin{subequations}
 \label{betaUSq3}
 \begin{align}
  \label{diff_betaa-3}
  \beta_{a-} &=\frac{\gamma_0 -\gamma_e -\upsilon+ i\Omega }{\gamma_0 +\gamma_e +\upsilon - i\Omega }\cdot \alpha_{ a-} +  \frac{2\sqrt{\gamma_0\gamma_e}\, \epsilon_{a-}}{\gamma +\upsilon - i\Omega}\\
  &  -\frac{4\gamma_0\, \eta^2 C_0^2}{(\gamma +\upsilon - i\Omega)^2(\gamma_m -i\Omega)} \left(\alpha_{a+} +\sqrt\frac{\gamma_e}{\gamma_0}\, \epsilon_{a+}\right)-\nonumber\\ 
	&-\frac{2\sqrt \gamma_0\,\eta C_0}{(\gamma +\upsilon - i\Omega)(\gamma_m -i\Omega)} \Big(\sqrt {2 \gamma_m} q_a +  f_{s\,a}\Big).
  \end{align}
  \end{subequations} 
and finally obtain equations \eqref{betaUSq} for sum and difference amplitude quadratures $\beta_{a\pm}$.


\end{document}